\documentclass[aps,prd,10pt,nofootinbib,twocolumn,eqsecnum,showpacs,showkeys,superscriptaddress,preprintnumbers,altaffilletter]{revtex4-1}

\usepackage{dcolumn}
\usepackage{amssymb}
\usepackage{amsmath}
\usepackage{amsfonts}
\usepackage{amsbsy}
\usepackage{color}
\usepackage{rotating}
\usepackage[english]{babel}
\usepackage{multirow}
\usepackage{float}
\usepackage{aas_macros}
\usepackage[normalem]{ulem}
\usepackage{hyperref}
\usepackage{cleveref}

\usepackage{standalone}
\usepackage{import}
\usepackage{tikz} 
\tikzset{
  pics/carc/.style args={#1:#2:#3}{
    code={
      \draw[pic actions] (#1:#3) arc(#1:#2:#3);
    }
  }
}

\newcolumntype{x}[1]{>{\centering\arraybackslash\hspace{0pt}}p{#1}}

\begin{document}

\title{Breaking the mass-sheet degeneracy \\ with gravitational wave interference in lensed events}

\date{\today}

\author{P. Cremonese}
\email{paolo.cremonese@usz.edu.pl}
\affiliation{Institute of Physics, University of Szczecin, Wielkopolska 15, 70-451 Szczecin, Poland}
\author{J. M. Ezquiaga}
\email{NASA Einstein fellow; ezquiaga@uchicago.edu}
\affiliation{Kavli Institute for Cosmological Physics and Enrico Fermi Institute, The University of Chicago, Chicago, IL 60637, USA}
\author{V. Salzano}
\email{vincenzo.salzano@usz.edu.pl}
\affiliation{Institute of Physics, University of Szczecin, Wielkopolska 15, 70-451 Szczecin, Poland}

\begin{abstract}
The mass-sheet degeneracy is a well-known problem in gravitational lensing which limits our capability to infer astrophysical lens properties or cosmological parameters from observations.
As the number of gravitational wave observations grows, detecting lensed events will become more likely, and to assess how the mass-sheet degeneracy may affect them is crucial. Here we study both analytically and numerically how the lensed waveforms are affected by the mass-sheet degeneracy computing the amplification factor from the diffraction integral.
In particular, we differentiate between the geometrical optics, wave optics and interference regimes, focusing on ground-based gravitational waves detectors. In agreement with expectations of gravitational lensing of electromagnetic radiation, we confirm how, in the geometrical optics scenario, the mass-sheet degeneracy cannot be broken with only one lensed image. However, we find that in the interference regime, and in part in the wave-optics regime, the mass-sheet degeneracy can be broken with only one lensed waveform thanks to the characteristic interference patterns of the signal. Finally, we quantify, through template matching, how well the mass-sheet degeneracy can be broken. We find that, within present GW detector sensitivities and considering signals as strong as those which have been detected so far, the mass-sheet degeneracy can lead to a $1\sigma$ uncertainty on the lens mass of $\sim 12\%$. With these values the MSD might still be a problematic issue. But in case of signals with higher signal-to-noise ratio, the uncertainty can drop to $\sim 2\%$, which is less than the current indeterminacy achieved by dynamical mass measurements. \end{abstract}



\maketitle

\section{Introduction}

The population of gravitational waves (GWs) events has grown considerably since the first ever detection in September of 2015 \cite{Abbott:2016blz}. The latest release, GWTC-2 \cite{Abbott:2020niy}, added 39 new events during the first half of the third observing run, O3a, to the 11 already detected in the previous runs, O1 and O2 \cite{LIGOScientific:2018mvr}. With new observatories coming online such as KAGRA \cite{Akutsu:2018axf} and IndIGO \cite{Unnikrishnan:2013qwa}, and the upgrade of the already existing ones, the number of detections is expected to grow faster and faster in the next years. With the growth of GWs events number, the first detection of gravitational lensing (GL) of GWs is getting ever closer. 

Accounting for the possibility of lensing would be fundamental in future observing runs. If lensing is not taken into consideration, it might lead to biased results in determining the source properties \cite{Cao:2014}, since lensing magnification changes the inferred luminosity distance and source masses. Moreover, in the limit of strong lensing, there could be also distortions in the lensed signal when higher modes, precession or eccentricity are relevant \cite{Ezquiaga:2020gdt}. Future analyses would require including lensed waveforms in the template bank when performing parameter estimation \cite{Nakamura:1997sw,Hannuksela:2019kle,Lai:2018rto,Pang:2020qow,Dai:2020tpj}, or adopting different strategies, as deep learning \cite{Singh:2018csp,Kim:2020xkm}. Lensing would also be relevant for population analyses since it would affect the inferred redshift and mass distributions, as the amplification of the signals allow to hear events further away.

For present detectors, though, the probability of a lensing event is pretty low. Such a probability will depend on both the mass of the lens and on the mass and distance of the source, but one can approximately state that, for binary black hole (BBH) systems, the rate of detection of strong lensing is $\approx 0.06$ yr$^{-1}$ for sources at redshifts $z<1$ \cite{Oguri:2019}. More in detail, in \cite{Ken:2018} it is shown that strong lensing rate of GWs produced by elliptical galaxies is $\approx 0.2$ yr$^{-1}$ for LIGO and $\approx 1$ yr$^{-1}$ for aLIGO, in accordance with \cite{Li:2018}. Moreover,  studies made on the existing data reject the possibility that observed GWs from BBH might be lensed by (known) clusters at $\leq 4\sigma$ \cite{Smith:2018}, as well as no evidence for any lensing event has been found for GW$190521$ \cite{Abbott:2020mjq}, although ``intriguing'' candidates might be already at our disposal \cite{Dai:2020tpj}.

Things will change as long as the GW events catalog becomes more numerous and detector's horizons expand. For example, in \cite{Li:2018} it is additionally stated that for the nominal Einstein Telescope sensitivity, the rate of lensing events could rise to $\approx 80$ yr$^{-1}$, while in \cite{Christian:2018vsi} it is discussed how third generation detectors will be able to enlarge the possible range of lensing events to lens masses as low as $1$ M$_{\odot}$.

Once a GW event is recognised as lensed, another problem, intrinsic to gravitational lensing, must be taken into consideration: the mass-sheet degeneracy (MSD) \cite{Falco85_MSD,Gorenstein88_MSD,Schneider:2013sxa,Schneider:2013wga}. The MSD relies on the fact that the simultaneous scaling of the lens mass and of the source plane leave the geometrical lensing outputs unchanged. Thus, in practice, we have an intrinsic indeterminacy when we analyze observational data to reconstruct the mass (model) of the lens from the images structure. This is a well-known problem in electromagnetic (EM) GL. For example, such degeneracy can affect the mass modelling of the lens objects and can highly bias the estimation of the Hubble constant, $H_0$ \cite{Schneider:2013sxa,Kochanek:2019ruu}, contributing to the systematic error budget. This is the reason why MSD is carefully analyzed in projects like H0LiCOW \cite{Chen:2019ejq,Wong:2019kwg}, TDCOSMO \cite{Millon:2019slk} and in the Time Delay Lens modelling Challenge \cite{Ding:2018hai,Ding:2020jmg}.

In EM studies, the MSD is always defined in the \textit{geometrical optics} (GO) approximation, when the wavelength of the lensed signal is much smaller than the typical size of the lens. In this limit,
the MSD can be broken, or at least attenuated, when: $1)$ multiple images of the source are available \cite{Lubini:2013mta,Grillo:2020yvj}; $2)$ the same gravitational object act as lens for multiple sources at different redshifts \cite{Saha:2000kn}; $3)$ independent mass estimations of the lens are available, e.g. from spatially resolved kinematical observations of the lens \cite{Schneider:2013sxa,Birrer:2020tax}; $4)$ by a proper combination of the previous astrophysical data with cosmological geometrical data \cite{Chen:2020knz}.

In the context of GL of GWs, given the much lower frequency of the signal, the \textit{wave optics} (WO) regime can also be relevant as GW wavelengths can be comparable to the lens size \cite{Takahashi:2016jom,Cremonese:2018cyg,Cremonese:2019tgb} and its features have been analyzed recently \cite{Jung:2017flg,Dai:2018enj,Sun:2019ztn,Liao:2020hnx}. In this limit, the MSD can be broken, at least in case of multiple lenses. In fact, the diffraction pattern is strictly connected to the geometrical configuration of multiple lenses \cite{Feldbrugge:2020ycp,Feldbrugge:2020tti}. Note that even in the EM GO regime, substructures in the lens may contribute to the uncertainty in the $H_0$ estimation \cite{Gilman:2020fie}.

Previous works have considered the possibility to infer the lens' masses from wave optics GW lensing in some specific conditions, see for example \cite{Takahashi:2003ix}. However, no explicit study is present in literature assessing the role of the MSD in GL of GWs. Here, for the first time, we take into consideration this problem. On the one hand we show how, in the classical GO case, also for GWs, the MSD still stands. On the other hand, when we are in a WO or WO-to-GO scenario (which we generally define as ``interference regime'') the GW waveform is sensitive to the MSD, exhibiting some intrinsic and characteristic features which thus allow us to break the MSD. It is important to stress that we find that, differently from the EM case \cite{Lubini:2013mta,Grillo:2020yvj,Chen:2019ejq}, the MSD can be solved even for a single observation of a single waveform, i.e. when no multiple images of the source are identified, and with one single lens object. 

This paper is organized as follows. In Sec.~\ref{ch:GL} we introduce the GL of GW, explaining the difference between GO and WO, and showing how the MSD affects the lensing of the waveforms. Sec.~\ref{ch:GL_regimes} will cover the calculations in GO, in the interference regime, and in WO. Here, we show when the MSD can be solved, with some examples. In Sec.~\ref{ch:SNR}, an additional quantitative analysis to assess this point and to define how well the MSD can be broken is presented. A discussion on the limits of the MSD breaking in WO is done and conclusions are drawn in Sec.~\ref{ch:conclusions}.

\section{Gravitational Lensing}\label{ch:GL}

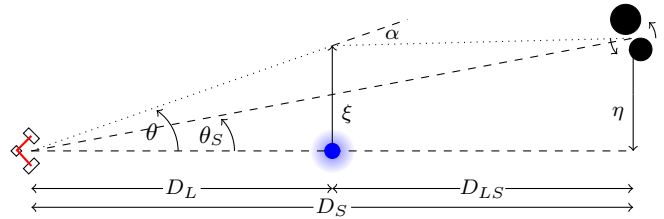
\begin{figure}[h]
\begin{tikzpicture}[scale=1,wave/.style={thick,color=#1,smooth}]

\tikzstyle{ann} = [font=\footnotesize,inner sep=1pt]
\draw[white,shading=radial,outer color=white,inner color=blue!50] (0,0) circle (0.3cm);


\draw[dashed] (-4,0) -- (4,0); 



\draw (-2.+0.1,0)  pic[black, ->]{carc=0:45:0.6cm}; \node at (-1.7+0.1,0.20) {$\theta_S$};
\draw (-3.+0.3,0)  pic[black, ->]{carc=0:56:0.65cm}; \node at (-2.7+0.3,0.25) {$\theta$}; 

\draw[<-] (-4,-0.75) -- (-.25,-0.75); 
\draw[] (0,-0.75) node[ann]{$D_S$};
\draw[->] (.25,-0.75) -- (4,-0.75);
\draw[<-] (0,-0.5) -- (2-0.3,-0.5);
\draw[] (2,-0.5) node[ann]{$D_{LS}$};
\draw[->] (2.3,-0.5) -- (4,-0.5);
\draw[<-] (-4,-0.5) -- (-2-0.25,-0.5);
\draw[] (-2,-0.5) node[ann]{$D_L$};
\draw[->] (-2.0+0.25,-0.5) -- (0,-0.5);
\draw[<-] (0,0) -- (0,.75);
\draw[] (0.2,.5) node[ann]{$\xi$};
\draw[->] (0,.75) -- (0,1.4);
\draw[<-] (4.,0) -- (4.,.75); 
\draw[] (3.8,.5) node[ann]{$\eta$};
\draw[->] (4.,.75) -- (4.,1.4);
\draw[rotate around ={45:(-4.2,0.)}] (-4.25,-.05) rectangle (-4.15,0.05);
\draw[rotate around ={-45:(-4.,0.2)}] (-4.1,0.15) rectangle (-3.9,0.25);
\draw[rotate around ={45:(-4.,-0.2)}] (-4.1,-0.25) rectangle (-3.9,-0.15);
\draw[thick,red] (-4.2,0) -- (-4.,0.2);
\draw[thick,red] (-4.19,0.01) -- (-4.,-0.2);

\draw[dashed] (-4,0) -- (4,1.5);

\draw[dotted] (-4,0) -- (0,1.4) -- (4,1.5);

\draw[dashed] (0,1.4) -- (1,1.75);
\draw[] (0.8,1.55) node[ann]{$\alpha$}; 

\draw[blue,fill=blue] (0,0) circle (0.1cm);

\draw[black,fill=black] (3.9,1.75) circle (0.2cm);
\draw[black,fill=black] (4.1,1.35) circle (0.15cm);
\draw (4.,1.5)  pic[black, ->]{carc=0:45:0.3cm};
\draw (4.,1.5)  pic[black, ->]{carc=180:180+45:0.3cm};


\end{tikzpicture}
\caption{Geometry of a gravitational lens system.}
\label{fig:lens_scheme}
\end{figure}

In this work we study the possibility to detect a GL event by some foreground object (lens) which alters the GW signal. The typical depiction of a gravitational lensing system is shown in \figurename~\ref{fig:lens_scheme}. Here, we have a BBH system as GW source, at an angular-diameter distance $D_S$ from the observer, and a lens, at distance $D_L$ from the observer and $D_{LS}$ from the source, which bends the trajectory of the GW and alters its waveform. The redshift locations of the source and the lens are $z_S$ and $z_L$, respectively. The position of the source and its image are described by the angles $\vec{\theta}_{S}$ and $\vec{\theta}$, respectively, while $\vec{\alpha}$ is the deflection angle. In this work, we always describe the lens with a point mass (PM) model. This is because we consider lenses that have $M_{L}\leq 10^4$~M$_\odot$. Within this mass range, and given the lens redshift which we work with, the most logic and realistic model is exactly the PM one.

Usually, the GL of light can be described and studied in the GO regime. For GW, though, we need a WO approach, because the corresponding conditions are more naturally satisfied. GO can be defined from the stationary phase approximation as the regime in which
\begin{equation} \label{eq:GO_cond}
    f\cdot\Delta t_d\gg1~,
\end{equation}
where $\Delta t_d$ is the time delay between the different stationary points (images) and $f$ is the observed frequency of the lensed radiation. From Eq.~\eqref{eq:GO_cond}, one can show that GO approximation breaks when \cite{Takahashi:2016jom}
\begin{equation} \label{eq:GO_cond2}
M_{L}\leq10^5 M_{\odot}\left[\frac{(1+z_L)f}{\text{Hz}}\right]^{-1},
\end{equation}
where $M_{L}$ is the (rest frame) mass of the lens. Below this threshold, WO is the right approach to described the lensing of the GW waveform.

In GO, the time delay is defined as
\begin{equation}\label{timedelGO}
    t(\vec{\theta},\vec{\theta}_{S})=\frac{1+z_L}{c}\frac{D_L D_S}{D_{LS}}\left[\frac{1}{2}(\vec{\theta}-\vec{\theta}_{S})^2-
    \hat{\Psi}(\vec{\theta})\right] ,
\end{equation}
with $\hat{\Psi}(\vec{\theta})$, the effective lens potential, being
\begin{equation}\label{lenspot}
\hat{\Psi}(\vec{\theta}) = \frac{D_{LS}}{D_L D_S}\frac{2}{c^2}\int \Phi(D_L\vec{\theta},z')dz'\, ,
\end{equation}
where $\Phi$ is the gravitational potential of the lens and $z'$ is the line-of-sight coordinate. An alternative way to write the previous expression in terms of dimensionless quantities is
\begin{equation}\label{timedelNODim}
t(\vec{x},\vec{y}) = \frac{1+z_L}{c} \frac{D_S \xi^{2}_0}{D_L D_{LS}} \left[ \frac{1}{2} \left( \vec{x}-\vec{y} \right)^2 - \Psi(\vec{x})\right]\; ,
\end{equation}
where: $\vec{\xi}_0$ is a reference scale length on the lens plane whose value depends on the mass model of the lens (for PM models, it is generally assumed to be the Einstein radius, $\xi_0 = D_L \theta_E$); $\vec{x}=D_L\vec{\theta}/\xi_0$ and $\vec{y}=D_L\vec{\eta}/D_S \vec{\xi}_0=D_L\vec{\theta}_{S}/\vec{\xi}_0$ are the dimensionless relative positions of the image and of the source (on the lens plane); $\vec{\xi}$ and $\vec{\eta}$ are the physical lengths on the lens and source plane, respectively; and $\Psi(\vec{x}) = D^{2}_L/\xi^{2}_{0}\hat{\Psi}(\vec{\theta})$ is the dimensionless lens potential. We can also define the dimensionless time delay as
\begin{equation}\label{eq:tGOdimless}
T(\vec{x}, \vec{y})=\frac{c}{1+z_L}\frac{D_L D_{LS}}{D_S \xi^{2}_0}\cdot t(\vec{x},\vec{y})\, .
\end{equation}
When we consider GL of GWs, if $h(t)$ is the GW strain (i.e. the waveform) in time domain, and $\tilde{h}(f)$ is the strain in the frequency domain, then the lensed waveform in the frequency domain will be given by
\begin{equation}
    \tilde{h}_{L}(f)=\tilde{h}(f)\cdot F(f,\vec{y}).
\end{equation}
Here, $F(f,\vec{y})$ is the amplification factor which comes from the diffraction integral.
In dimensionless units this is defined as \cite{gralen.boo}
\begin{equation}\label{eq:ampfactor}
F(w,\vec{y})=\frac{w}{2\pi i}\int \mathrm{d}^2x\exp[iwT(\vec{x},\vec{y})]\; ,
\end{equation}
where $w=\frac{1+z_L}{c}\frac{D_S \xi^{2}_{0}}{D_LD_{LS}} 2\pi f$ is the dimensionless frequency of the GW. Assuming spherical symmetry \cite{Nakamura7}, we can express it as
\begin{eqnarray} \label{eq:AF}
F(w,y) &=& -iw e^{iwy^2/2} \\
&\times& \int_0^\infty \mathrm{d}x\,x\,J_0(wxy)\exp\left\{iw\left[ \frac{1}{2}x^2-\Psi(x)\right]\right\}\,, \nonumber
\end{eqnarray}
where $J_0$ is the Bessel function of zeroth order, and, since we are assuming spherical symmetry, we can write $x=|\vec{x}|$ and $y=|\vec{y}|$.
In simple terms, the amplification factor accounts for all possible trajectories of the signal around the lens. Independent images are formed when the stationary points are enough separated in time (i.e. in GO regime), leading to an amplification factor of the form
\begin{equation}\label{Fgeom}
    F\approx \sum_j\vert\mu(\vec\theta_j)\vert^{1/2}\exp\left(i\omega t_d(\vec\theta_j)-i\, \text{sign}(\omega)\frac{n_j\pi}{2}\right),
\end{equation}
where $\mu(\vec\theta_j)$ is the magnification of the $j$-th image located at $\vec\theta_j$, and $n_j$ is a frequency-independent phase shift associated with the type of image or extrema of the time delay surface ($n_j=0$ for type I, $n_j=1$ for type II, and $n_j=2$ for type III images, see \cite{gralen.boo} for details).

\subsection{Mass-sheet degeneracy}\label{ch:MSD}

The MSD is based on the following set of transformations which leave the lensed signal observables unchanged \cite{Falco85_MSD,Gorenstein88_MSD}:
\begin{enumerate}
 \item scaling by the same factor $\lambda$ both $\vec{\alpha}$, the reduced angle of deflection, and $\vec{\theta}_{S}$, the source position (angle):
\begin{equation}\label{eq:msd1}
\vec{\alpha} \rightarrow \vec{\alpha}_{\lambda} = \lambda \vec{\alpha} + (1-\lambda) \vec{\theta}\,,
\end{equation}
\begin{equation}\label{eq:msd2}
\vec{\theta}_{S} \rightarrow \vec{\theta}_{S,\lambda} = \lambda \vec{\theta}_{S}\,.
\end{equation}
\item scaling the lens mass (expressed in terms of the convergence $\kappa$ \cite{gralen.boo}) by adding or subtracting a constant mass layer (sheet) which acts as an additional thin lens:
 \begin{equation}\label{eq:msd3}
\kappa \rightarrow \kappa_{\lambda} = \lambda \kappa + (1-\lambda)\,.
\end{equation}
\end{enumerate}
The redshifts and the intrinsic source luminosity are not affected by MSD transformations. 
Note that the parameter $\lambda$ is nowadays more commonly recognized to technically take into account two different contributions \cite{Birrer:2016xku}: not only mass from the lens itself (generally called ``internal'' MSD), but also environmental mass effects, from the matter distributed out of the lens and projected along the line of sight, thus called ``external'' MSD.

Without going into all the details and consequences of the above transformations, which have been thoroughly studied \cite{Falco85_MSD,Gorenstein88_MSD}, we focus here on the main features which are relevant for our study. In particular, the dimensionless source position, the lens potential and the time delay can be found to change under MSD transformations as
\begin{equation}
\vec{y} \rightarrow \vec{y}_{\lambda} = \lambda \vec{y}\, ; \label{eq:yT}
\end{equation}
\begin{equation}
\Psi(\vec{x})  \rightarrow \Psi_{\lambda}(\vec{x}) = \lambda \Psi(\vec{x}) + (1-\lambda) \frac{{|\vec{x}|^2}}{2}\, ; \label{eq:potT}\\
\end{equation}
\begin{equation}
t(\vec{x},\vec{y}) \rightarrow t_{\lambda} = \lambda\, t
- \frac{\lambda(1-\lambda)}{2} \left( \frac{1+z_L}{c}\frac{D_S \xi^{2}_0}{D_{L} D_{LS}}\right) \vec{|y|}^{2}\, . \label{eq:tdelT}
\end{equation}
Therefore, the amplification factor needs to be modified. Starting from Eq.~(\ref{eq:AF}), inserting the right transformed quantities Eqs.~(\ref{eq:yT})~-~(\ref{eq:potT})~-~(\ref{eq:tdelT}), and always relying on spherical symmetry, we get:
\begin{align} \label{eq:AFtransf}
F_{\lambda}(w,y) &= -iw e^{iw \lambda^2 y^2/2} \\
&\cdot \int_0^\infty \mathrm{d}x\,x\,J_0(\lambda wxy)\exp\left\{iw \lambda \left[ \frac{x^2}{2}-\Psi(x)\right]\right\}\,. \nonumber
\end{align}

Since, as we have stated above, we are working only with PM lenses, we can solve this integral analytically, getting:
\begin{align}\label{eq:AFtransf_A}
    F_{\lambda}(w,y)
    &= \frac{1}{\lambda} \left(-\frac{i \lambda  w}{2}\right)^{1+\frac{i \lambda  w}{2}} \exp\left\{\frac{1}{2} i w (\lambda y)^2\right\} \,\\
   &\cdot \Gamma
   \left(-\frac{i w \lambda}{2} \right)\ _1F_1\left(1-\frac{i w \lambda }{2};1;-\frac{i w \lambda}{2} y^2 \right)~, \nonumber
\end{align}
where $\Gamma$ is the gamma function and $_1F_1$ is the confluent hypergeometric function. Note that in the limit $\lambda \rightarrow 1$ we get the standard result shown in \cite{Peters:1974gj,Takahashi:2003ix}. It is also to be noted that for PM, there will be only two images in the strong lensing limit. We will denote them as + and - referring to their magnification and de-magnification respectively.

In the next sections, we will use this modified amplification factor (\ref{eq:AFtransf_A}) to compute the lensing of GWs under MSD. Although in general it is difficult to guess analytically the effect of the MSD through $\lambda$, this becomes more apparent in the GO. There one can see that $\lambda$ will only appear in the exponential of (\ref{Fgeom}) attached to the time delay. Therefore, we can anticipate that $\lambda$ will not introduce any characteristic distortion of the independent images, apart for a magnification effect, that would allow us to break the MSD.

\section{Optical Lensing regimes}\label{ch:GL_regimes}

\begin{figure*}[htb!]
\centering
\includegraphics[width=0.9\textwidth]{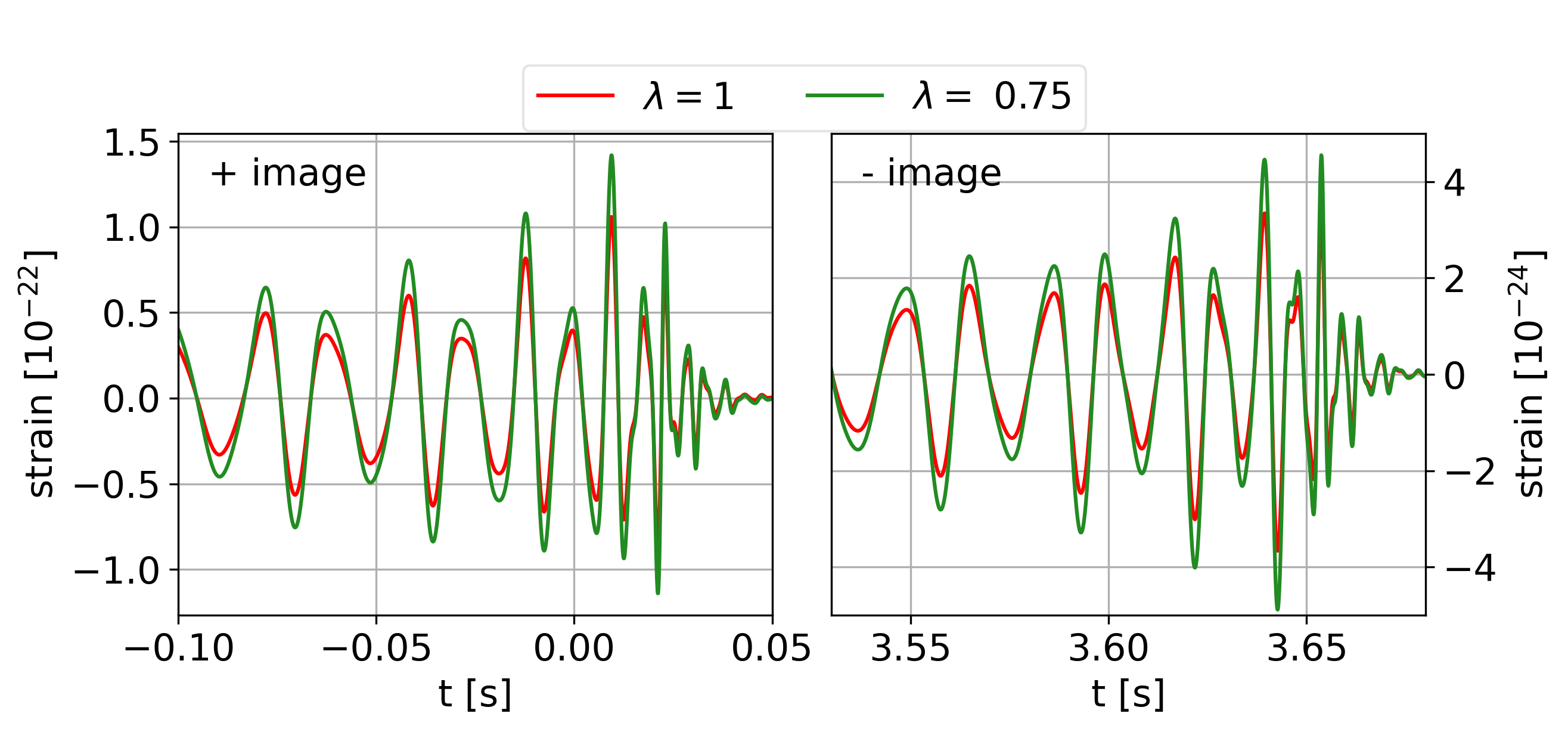}
\caption{Impact of the MSD transformations on the time domain waveforms of lensed GWs in the geometrical optics limit. We choose a lens with (rest) mass $M_{L}=10^4 M_\odot$ and at redshift $z_{L}=0.1$. The source has a (rest) total mass of $M_{\rm tot} = 60 M_\odot$ at redshift $z_S=0.5$ and $y=5$ with $q=0.1$, and $\left\{s_{1,z},s_{2,z}\right\}=\left\{0.7,0.2\right\}$. In each panel we show, after arbitrary time alignment, the waveform after MSD with $\lambda=0.75$ in green and the original one in red. We focus on the peaks of the positively magnified (left) and negatively magnified (right) images. The difference in the y-axis of each image accounts for the effect of (de-)magnification.}
\label{fig:MSD_go_s0702}
\end{figure*}

Now that we have defined how the MSD affects the lensing of GWs, we will show how it can be broken in some cases. The chosen procedure will be to compute the lensed waveforms for different values of the MSD parameter, in particular $\lambda=\{0.5,0.75,0.95,1.1\}$, and to compare them to the \textit{original} signal, i.e. when $\lambda=1$, with respect to which we aim to solve the degeneracy. It is important to stress here that such \textit{original} signal is the waveform that would be detected by a possible instrument, and whose analysis, if the MSD is not properly taken into account, might lead to biased estimations of the lens and source systems. For the sake of clarity and compactness, we are going to show only some selected explanatory examples to clarify the content of our statements. For all cases, we fixed the values of the inclination of the source w.r.t. the line of sight at $\iota=\pi/3$, and the sky position and orientation w.r.t. the detector \cite{Sathyaprakash:2009xs} at $\theta_d=0.3$, $\phi_d=0.4$ and $\Psi_d=1.5$.

\subsection{Geometrical optics regime}\label{sch:GO}

To show how MSD acts in the GO case and to crosscheck our pipeline, we first consider, as source of the GW signal, a BBH merger with a (rest) total mass of $M_{\rm tot} = 60$ M$_\odot$ at redshift $z_S=0.5$, and three different scenarios: \textit{a)} same BH masses, $q=m_2/m_1=1$, with $m_2\leq m_1$; \textit{b)} different BH masses, $q = 0.1$; \textit{c)} different BH masses, $q = 0.1$, and (aligned) spins $\{s_{1,z},~s_{2,z}\}=\{0.7,~0.2\}$. The lens has a (rest) mass $M_{L}=10^4\,M_\odot$ and a redshift $z_{L}=0.1$. Since we focus on PM lenses, the time delay between the two images is uniquely determined by the source position $y$ (see e.g. Eq. (18) in \cite{Ezquiaga:2020spg}). For this concrete system, when the source position is $y=5$, we are safely in GO.\footnote{Note though that at large $y$ one approaches the regime of weak lensing where the first image is weakly magnified, $\mu_+\simeq 1 + 1/y^4$, and the second image highly de-magnified, $\mu_-\simeq 1/y^4$.} In fact, if we take into consideration the initial detected frequency, which in this case is $f_i = 30$ Hz, we can see how the conditions from Eqs.~\eqref{eq:GO_cond} and \eqref{eq:GO_cond2} become $f_i\cdot\Delta t_d \approx 10^2 \gg 1$ ($\Delta t_d\approx3.6$ s) and $M_L = 10^4M_\odot > 3\cdot10^3 M_\odot= 10^5 \left[(1+z_L)f_i\right]^{-1}M_\odot$. Thus, one can easily understand that, for greater frequencies, which are achieved while approaching the merging of the BHs, these conditions are even more strongly satisfied.

The unlensed waveform in time domain, $h(t)$, is obtained using the \texttt{PyCBC} \cite{alex_nitz_2020_4134752} software in \textsc{python}, with the \texttt{IMRPhenomHM} approximant \cite{London:2018} which includes higher-modes for circular, (anti)-aligned spinning sources. We evaluate the amplification factor, Eq.~\eqref{eq:AFtransf_A}, varying the transformation parameter $\lambda=\{0.5,0.75,0.95,1.1\}$ using \textsc{python}.  Then, we use the software \texttt{FFTW}\footnote{http://www.fftw.org/} to compute both the Fourier transform, i.e. to get the unlensed $h(f)$, and the inverse Fourier transform to get the lensed waveform $\tilde{h}_{L}(t)$.

In the simplest case, $q=1$, we verify how the observables are properly transformed under the MSD transformations. In particular, we define $p^{i}_{j}$ so that, for example, $p^{+}_{0.75}$ refers to the positively magnified maximum peak of the waveform with $\lambda= 0.75$. We identify the peak as the strain value of the positive maximum of the waveform; since we are in GO and no modulation is present (which is instead a typical feature of the WO regime), this is a safe definition. Then, we also calculate the time delay between the maximum peaks of the two image. We use the notation $\Delta t_{i}$ to indicate, for example, that $\Delta t_{0.75}$ is the ratio between the time delay difference with $\lambda=0.75$ with respect to the time delay difference with $\lambda = 1$.

In Table~\ref{tab:MSD_values} we present results of this first calculations. We can see how, as expected \cite{Falco85_MSD,Gorenstein88_MSD}: the ratio of the peaks/magnifications, for each separated image, with respect to the original waveform, scales as $\lambda$ (columns $2$ and $3$); the arrival time differences scale as $\lambda$ (column $4$).

\setlength\extrarowheight{5pt}
\begin{table}[htb!]
\centering
\begin{tabular}{c|c|c|c}
\toprule
\multirow{2}{*}{$\lambda$}	&	$p_1/p_\lambda$	& $p_1/p_\lambda$	&	\multirow{2}{*}{$\Delta t_\lambda$}	\\
 & $+$ im. & $-$ im. & \\
\hline
\hline
0.5  &	0.500	&	0.505	&	0.500 \\ 
0.75 &	0.750	&	0.754	&	0.750 \\ 
0.95 &	0.950	&	0.951	&	0.950 \\ 
1  	 &	1   	&	1   	&	1     \\
1.1	 &	1.101	&	1.097	&	1.100 \\ 
\hline
\end{tabular}
\caption{Scaling of observable under MSD transformations in geometrical optics. The source has a (rest) total mass of $M_{\rm tot} = 60 M_\odot$, at redshift $z_S=0.5$ and $y=5$ with $q=1$. The lens has a (rest) mass of $M_{L}=10^4 M_\odot$, and it is at redshift $z_{L}=0.1$. }
\label{tab:MSD_values}
\end{table}

Overall, the MSD is not solved in GO. The same conclusion can be visually derived also from Fig.~\ref{fig:MSD_go_s0702}, where we only show the case with mass asymmetry and spin for $\lambda = 0.75$, but results apply for all $\lambda$ values and in all three scenarios stated at the beginning of the section.

It is to be noted that, even in this scenario where the waveform has characteristic features due to the presence of higher order modes ($q=0.1$ and $\left\{s_{1,z},s_{2,z}\right\}=\left\{0.7,0.2\right\}$), which should ease the visualization of possible differences between the original and the transformed waveform, we cannot visually detect any sign of distortion induced by the MSD. The MSD only changes the magnification. If we look at the negatively magnified image in the right panel of Fig.~\ref{fig:MSD_go_s0702}, we also see no evidence at all for any kind of distortion.
Indeed, it is in type II images, where, if any, we would expect to find a characteristic feature dependent on the MSD due to the distortions with respect to general relativity (unlensed) waveforms due to the lensing phase shift when higher order modes are dominant \cite{Ezquiaga:2020gdt}.
As we can see, though, no such feature is present and the MSD cannot be broken. This visual analysis will be quantified precisely in Sec. \ref{ch:SNR} when considering the effect in the detector signal-to-noise.

\subsection{Interference regime}\label{ch:WO}

Since we did not found any distinctive signature of the MSD transformation in the GO limit, in this section, we analyze how the MSD affects the waveform when we are in the intermediate regime, transitioning from GO to WO, where most of the interference features should appear since the time delay between the images is of the order of the signal duration.

\begin{figure}[htb!]
\centering
\includegraphics[width=0.45\textwidth]{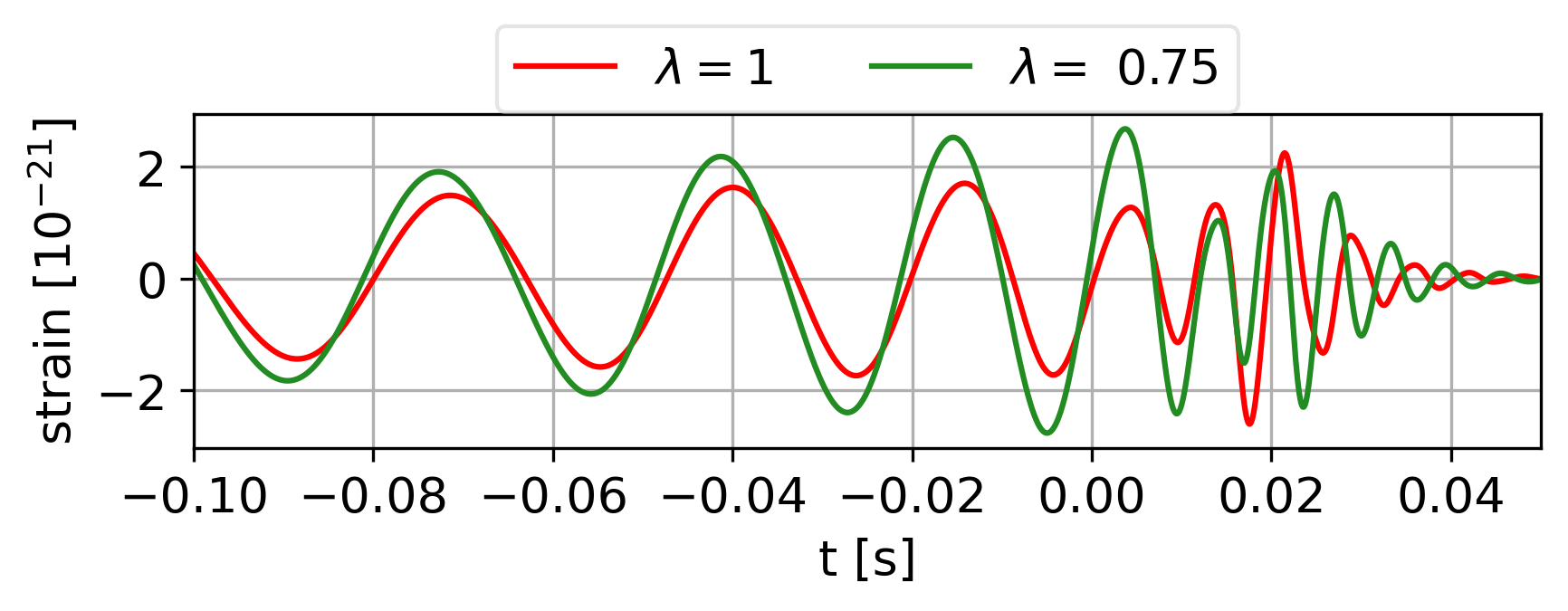}\\
\includegraphics[width=0.45\textwidth]{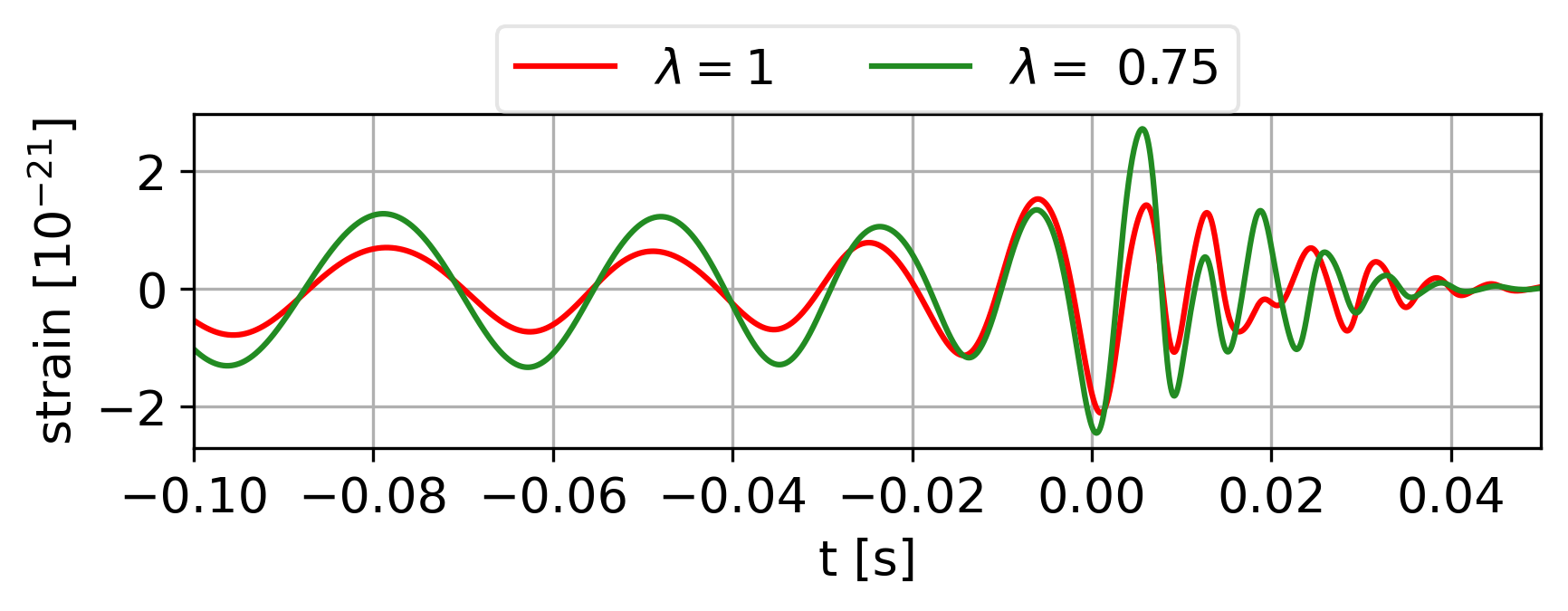}\\
\includegraphics[width=0.45\textwidth]{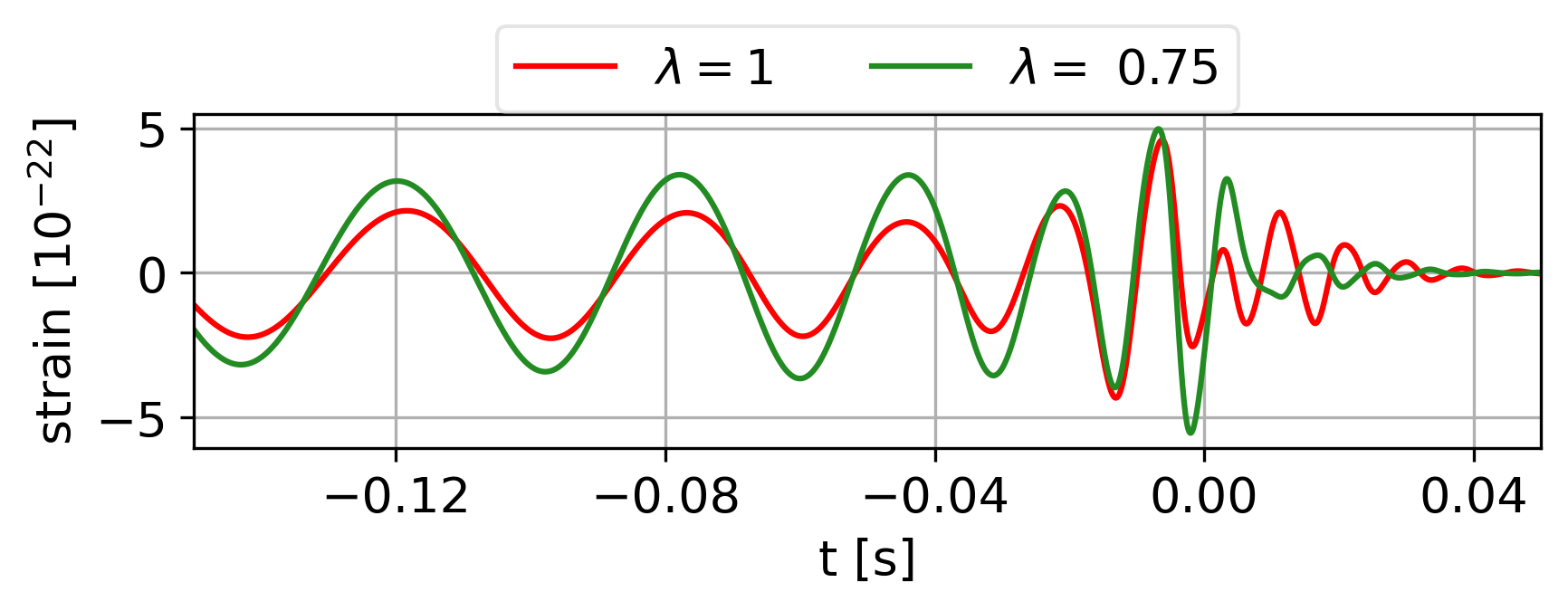}\\
\caption{Impact of the MSD transformations on the time domain waveforms of lensed GWs in the interference regime, time domain. Lens with (rest) mass $M_{L}=500~M_\odot$ and at redshift $z_{L}=0.01$. Source with a (rest) total mass of $M_{\rm tot} = 100~M_\odot$ and: $z_S=0.1$, $y=0.5$ in the top panel; $z_S=0.1$, $y=1$ in the middle panel; $z_S=0.5$, $y=1$ in the bottom panel. All sources have $q=1$. After arbitrary time alignment, we show the waveform after MSD with $\lambda=0.75$ in green and the original one, i.e. $\lambda=1$, in red.}
\label{fig:MSD_INT_TD_1}
\end{figure}

As source of our GW signal, we consider a BBH merger with a (rest) total mass of $M_{tot} = 100$ M$_\odot$ at redshift $z_S=0.1$, and the three different scenarios for source masses and spin considered before. The lens now has a (rest) mass $M_L=500$ M$_\odot$ and a redshift $z_L=0.01$. We consider a PM model as mass distribution for the lens and we focus on two source positions on the lens plane, namely $y=1$ and $y=0.5$. A third case we consider consists of the same lens and the same source mass positioned at $y=1$, but $z_S=0.5$. The tools used here to compute the waveforms and the amplification factor are the same as in the previous section.

\begin{figure}[htb!]
\centering
\includegraphics[width=0.45\textwidth]{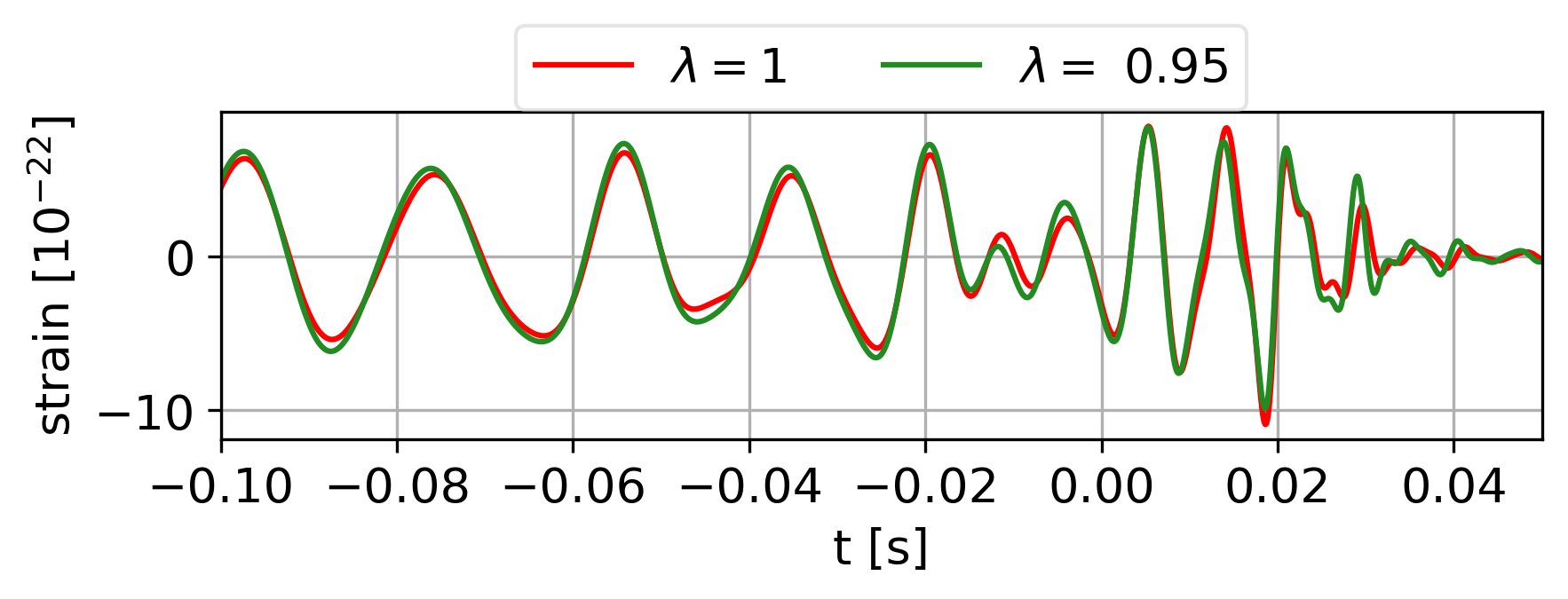}\\
\includegraphics[width=0.45\textwidth]{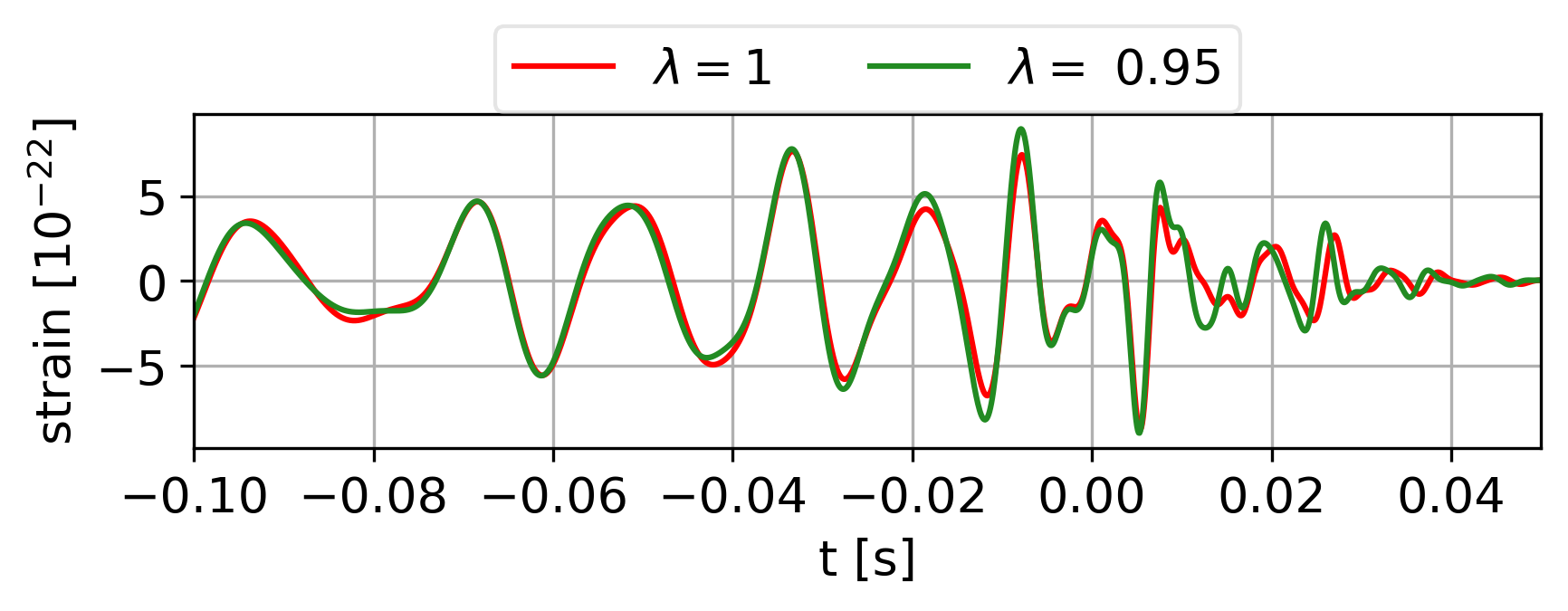}\\
\includegraphics[width=0.45\textwidth]{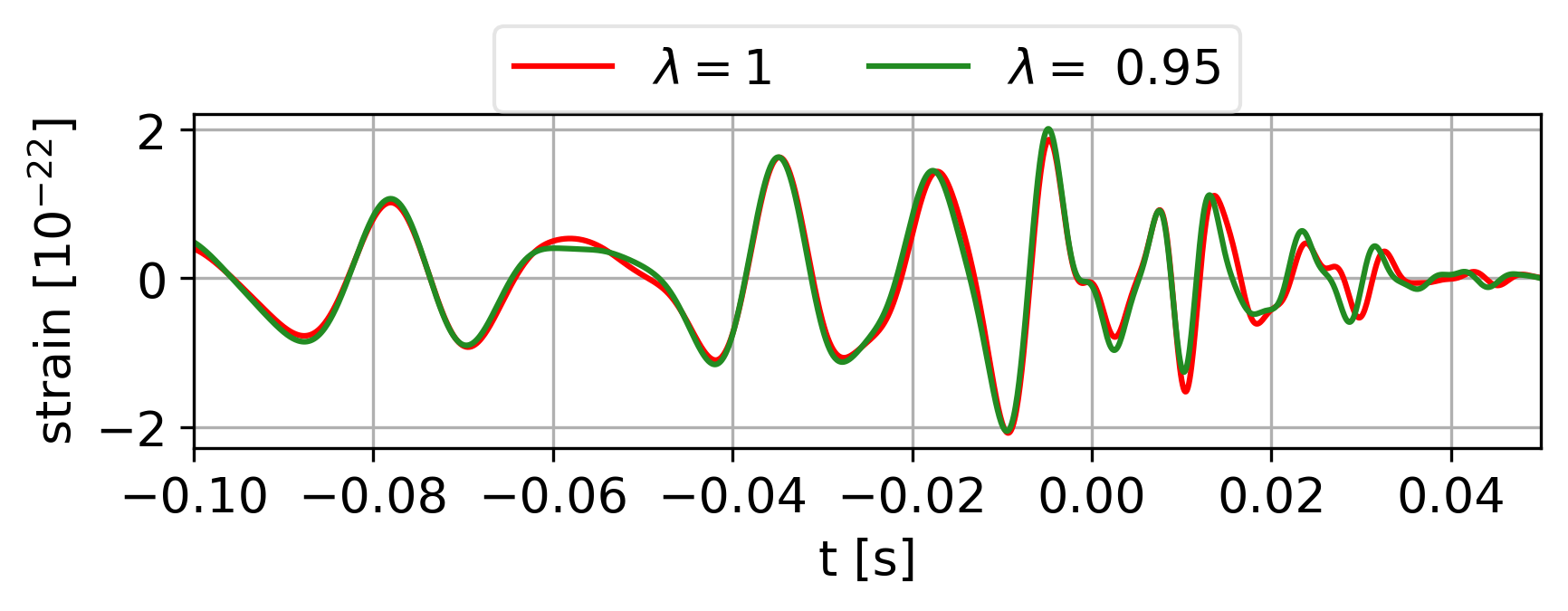}\\
\caption{Impact of the MSD transformations on the time domain waveforms of lensed GWs in the interference regime, time domain. Lens with (rest) mass $M_{L}=500~M_\odot$ and at redshift $z_{L}=0.01$. Source with a (rest) total mass of $M_{\rm tot} = 100~M_\odot$ and: $z_S=0.1$, $y=0.5$ in the top panel; $z_S=0.1$, $y=1$ in the middle panel; $z_S=0.5$, $y=1$ in the bottom panel. All sources have $q=0.1$ and $\left\{s_{1,z},s_{2,z}\right\}=\left\{0.7,0.2\right\}$. After arbitrary time alignment, we show the waveform after MSD with $\lambda=0.95$ in green and the original one, i.e. $\lambda=1$, in red.}
\label{fig:MSD_INT_TD_2}
\end{figure}

\begin{figure*}[htb!]
\centering
\includegraphics[width=0.32\textwidth]{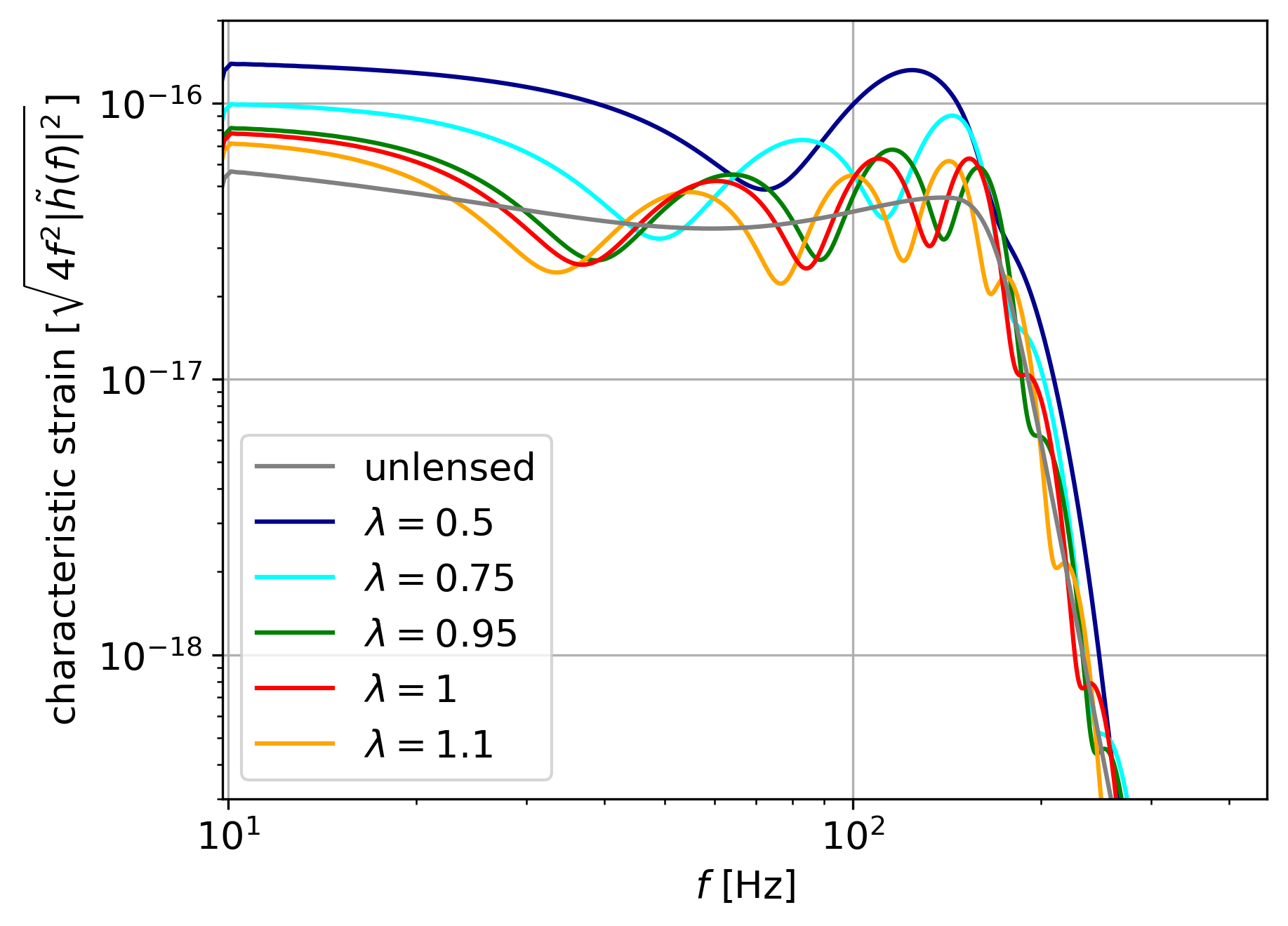}~
\includegraphics[width=0.32\textwidth]{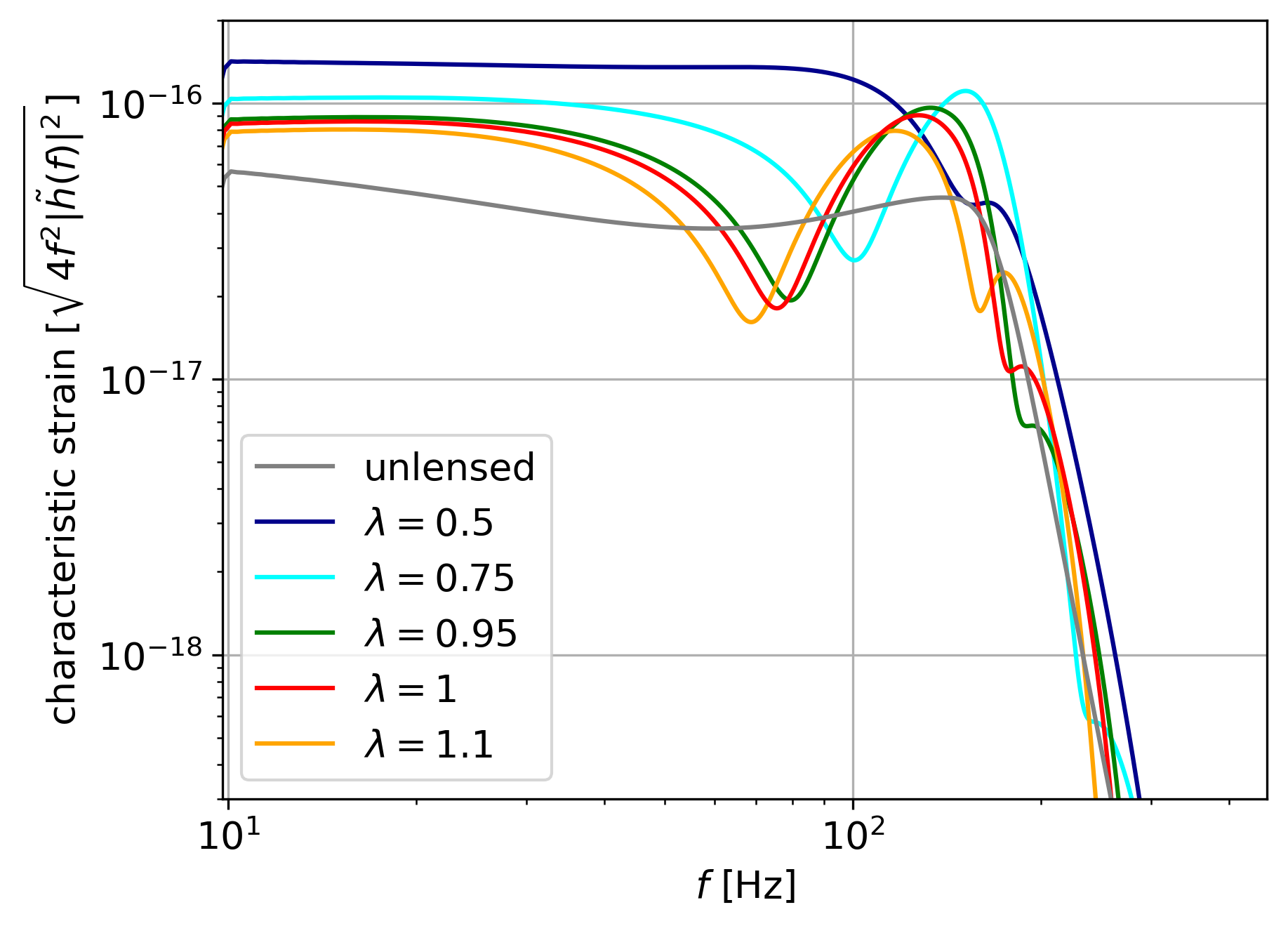}~
\includegraphics[width=0.32\textwidth]{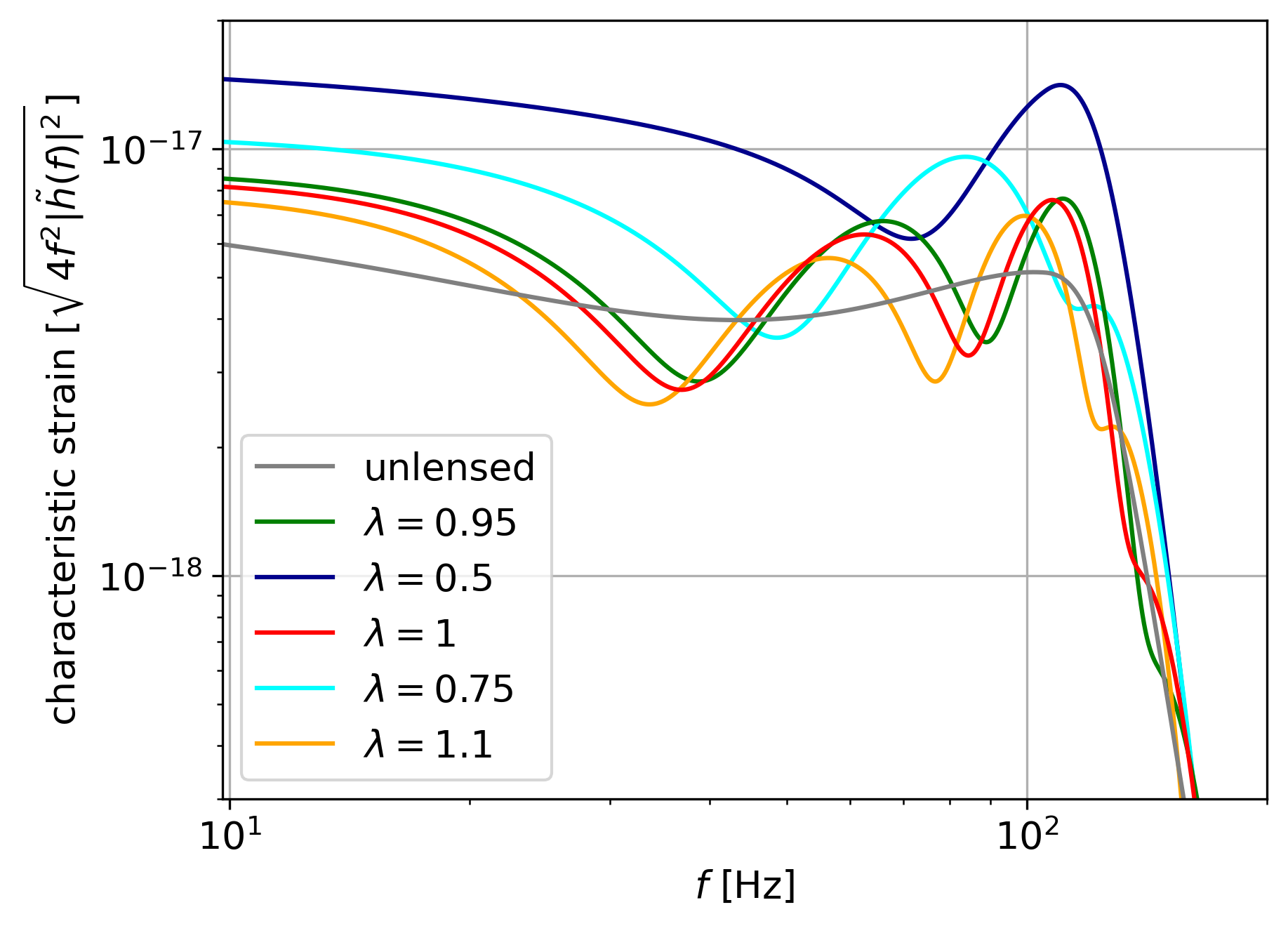}
\caption{Interference regime, frequency domain. Lens with (rest) mass $M_{L}=500 M_\odot$ and at redshift $z_{L}=0.01$. Source with a (rest) total mass of $M_{\rm tot} = 100 M_\odot$ at redshift $z_S=0.1$ and $y=1$ (left panel) and $y=0.5$ (central panel), and source at $z_S=0.5$ with $y=1$ (rights panel). Here, $q=1$ always. Comparing cases with different values for the MSD parameter $\lambda$.}
\label{fig:MSD_INT_FD}
\end{figure*}

We find that the lensed waveforms with different $\lambda$ are significantly and intrinsically different with respect to the case with $\lambda=1$, which is the original signal. As an example, in Figs.~\ref{fig:MSD_INT_TD_1} and \ref{fig:MSD_INT_TD_2} we show, respectively, how the waveform is changed when $\lambda=0.75$ and $0.95$, respectively. Additionally, in Fig.~\ref{fig:MSD_INT_FD} we show what happens in the frequency domain for all the MSD parameters $\lambda$ for the $q=1$ scenario.

In all three plots, we can see how the interference regime is different from the GO one. In fact, now, the transformation acts on the waveform not only on the overall amplitude of the strain, i.e. magnifying positively or negatively the image, but also on its shape. Let us consider Fig.~\ref{fig:MSD_INT_TD_1}, where we have $\lambda=0.75$ and we show the source (always with $q=1$) with $z_S=0.1$ and $y=0.5$ in the top panel, with $z_S=0.1$ and $y=1$ in the middle panel, and $z_S=0.5$ and $y=1$ in the bottom panel. In the top panel, we can see how for $t<0$ one has mostly a modulated magnification of the waveform (any temporal shift is basically undetectable) which might lead to think that there is not such a big difference w.r.t the GO regime. For $t>0$, though, we see how the pattern is strongly changing, with an out-of-phase magnification, and a distortion of the waveform for the case $\lambda=0.75$ which has no correspondence in the original signal. These characteristic features are even more evident in the central panel, where we can clearly distinguish a (positive) maximum when $\lambda=0.75$, not present in the original (red) waveform. Then, mostly at $t\sim 0.01-0.02$ we can also see that, when $\lambda=1$, a peculiar feature is present, caused by destructive interference, differently from $\lambda=0.75$, where it is much less visible. Finally, in the bottom panel, we can see how the two waveforms are totally different for $t>0$.

In general, from the plots we can deduce that, when $\lambda<0.95$, the distinction between the waveforms is more evident. But still, with $\lambda = 0.95$, we can detect some differences, as shown in Figs.~\ref{fig:MSD_INT_TD_2} and \ref{fig:MSD_INT_FD}. In Fig.~\ref{fig:MSD_INT_TD_2}, the panels are divided as in Fig.~\ref{fig:MSD_INT_TD_1}, but we are now considering $\lambda=0.95$ and the source has $q=0.1$ and $\left\{s_{1,z},s_{2,z}\right\}=\left\{0.7,0.2\right\}$. Here, the differences are more subtle. On the top panel, we can see how the modulation of the two waveforms starts to be different at $t\sim-0.01$, with some smaller further shape difference at $t\sim 0.03$. The same can be observed in the middle panel, where the two waveforms are intrinsically different at $t\sim0.01-0.02$, and in the bottom one.

If we consider the corresponding frequency domain, in all panels  of Fig.~\ref{fig:MSD_INT_FD}, the differences are straightforward to identify. Every waveform has a different shape, and we can not move from one to another with a simple rescaling, which would be characteristic of a simple magnification, which implies that they are intrinsically different. Note that in Fig.~\ref{fig:MSD_INT_FD} we only show the case $q=1$, but the same conclusions hold also for the other scenarios, that we are not showing here just for the sake of clarity, because when higher order modes are present, the waveforms in frequency domain become more messy.

Of course, one would expect that the smaller the changes (with $\lambda$), the more difficult it is to break the degeneracy. In Sec.~\ref{ch:SNR}, we will quantify more precisely such differences with a match-filter analysis.
However, it seems clear that, in principle, one single lensed waveform in the interference regime can break the MSD. The reason for that is given by the fact that the MSD transformation changes the time delay between the images and we are here very close to the boundary between WO and GO. In fact, since the two images, well separated in the GO case, here are interfering, a change in the time difference is translated into a different interference pattern.

\begin{figure*}[htb!]
\centering
\begin{minipage}[c]{.48\linewidth}
{\includegraphics[width=\textwidth]{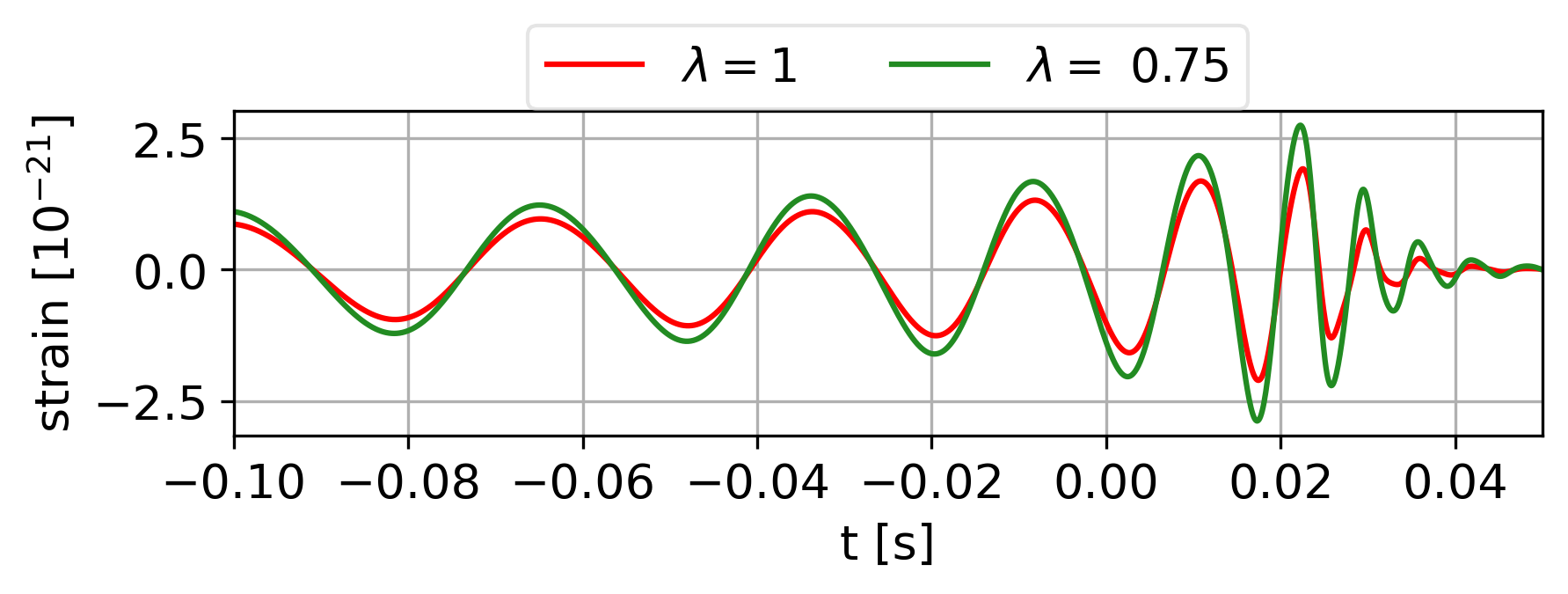}}\\
{\includegraphics[width=\textwidth]{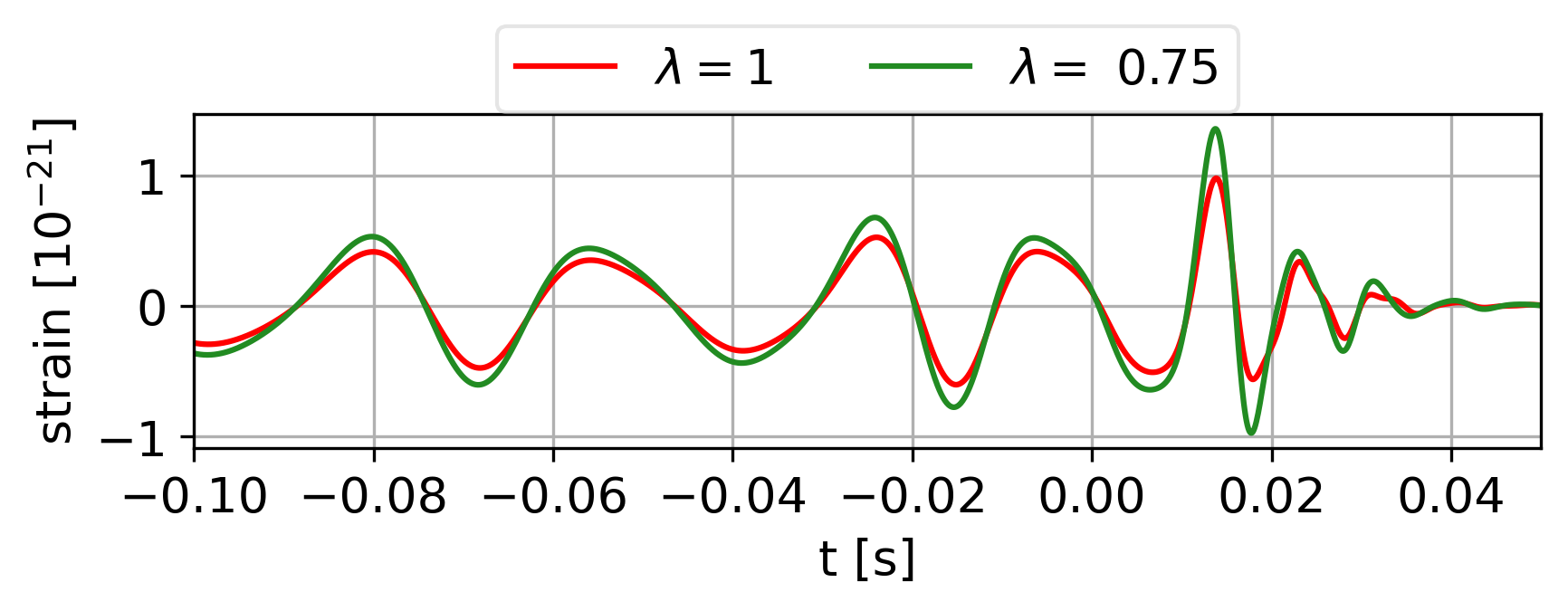}}\\
{\includegraphics[width=\textwidth]{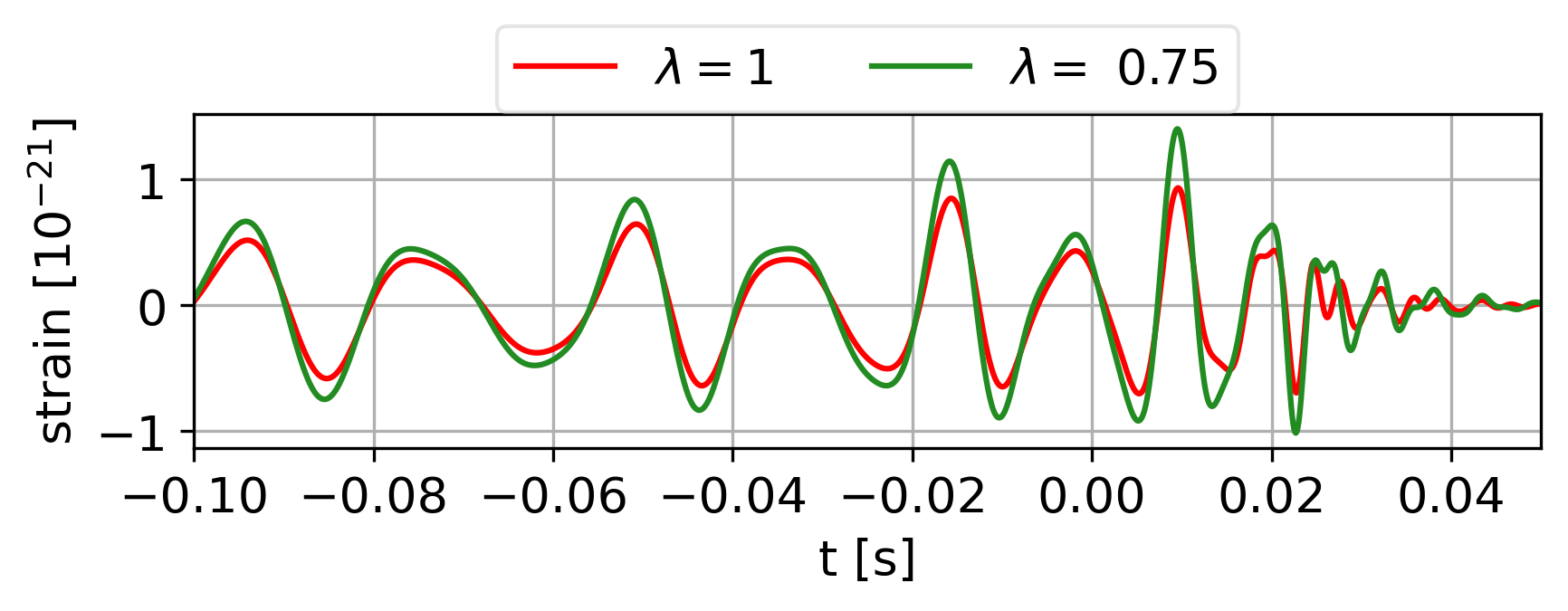}}
\end{minipage}
\hfill
\centering
\begin{minipage}[c]{.48\linewidth}
{\includegraphics[width=\textwidth]{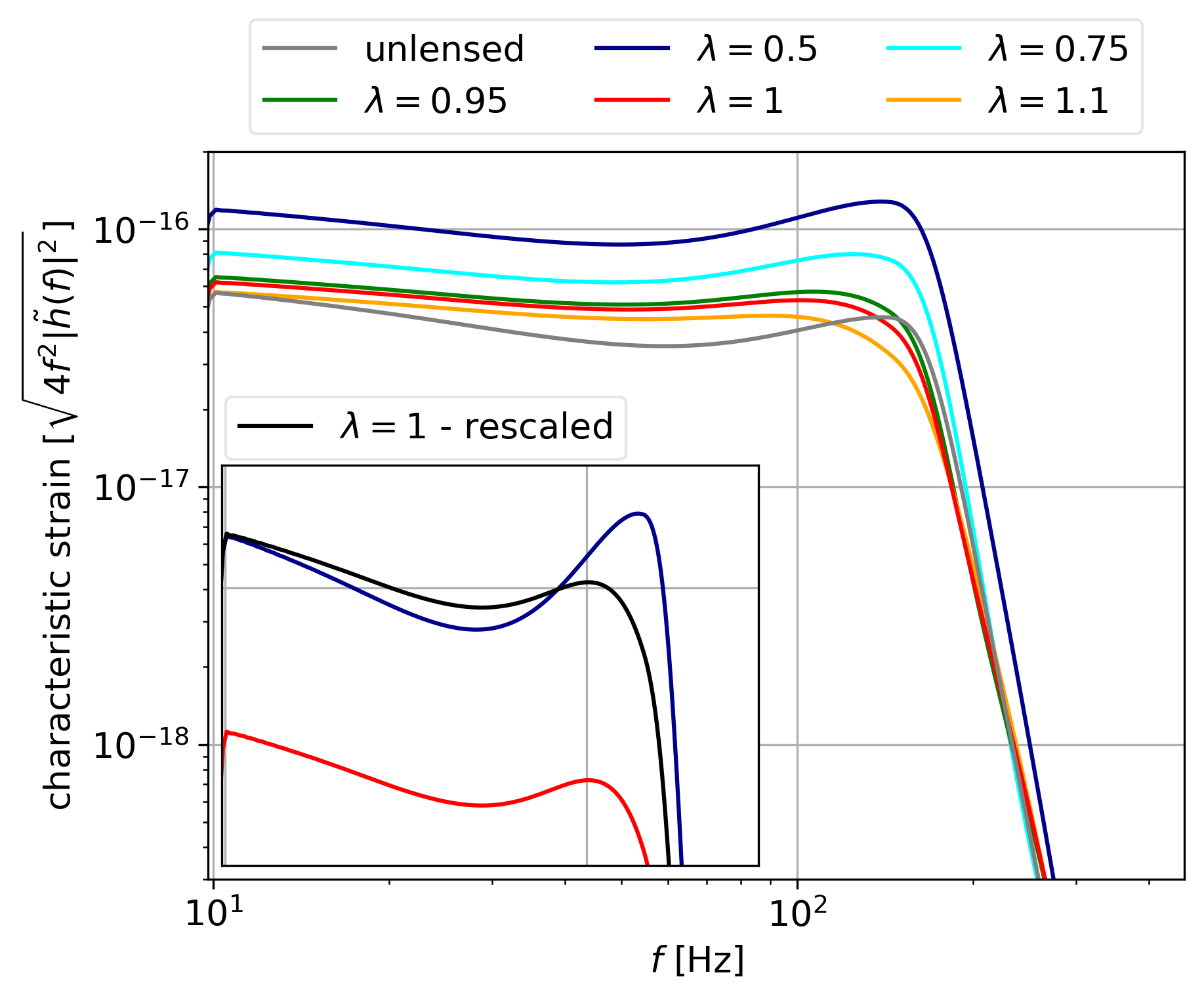}}
\end{minipage}
\caption{Wave optics regime, time domain (left panel) and frequency domain (right panel). Lens with (rest) mass $M_{L}=100 M_\odot$ and at redshift $z_{L}=0.01$. Source with a (rest) total mass of $M_{\rm tot} = 100 M_\odot$ at redshift $z_S=0.1$ and $y=1$. In the time domain, we show the cases for $\lambda=0.75$ and: $q=1$ (top panel); $q=0.1$ (middle panel); $\left\{s_{1,z},s_{2,z}\right\}=\left\{0.7,0.2\right\}$ (bottom panel). In the frequency domain, we compare cases with different values for the MSD parameter $\lambda$ for the $q=1$ scenario only. In the inset, we show a comparison between $\lambda=0.5$, $\lambda=1$ and a rescaled version of $\lambda=1$. More details in the text.}
\label{fig:MSD_WO}
\end{figure*}

\subsection{Wave optics regime}

Finally, we want to understand what happens when we get further into WO. Here, the possible breaking of the MSD is not always so clear. Actually, one should note that in the case of pure WO, i.e. when $w\rightarrow 0$, from Eq.~\eqref{eq:AFtransf_A} we get $F\to1$\footnote{Note that $F_{\lambda}$ is totally equivalent to Eq.~(\ref{eq:AF}) when expressed in terms of $y_{\lambda}$ and $\Psi_{\lambda}$. In that case, $F_{\lambda}\rightarrow 1$ for $w \rightarrow 0$. But our Eq.~\eqref{eq:AFtransf_A} is in terms of $y$ and $\Psi$ which correspond to $\lambda=1$. Thus, in Eq.~\eqref{eq:AFtransf_A}, we must perform both the limits $w \rightarrow 0$ and $\lambda \rightarrow 1$ in order to recover $F_{\lambda} \rightarrow 1$.} . Since no lensing is present, the MSD cannot be assessed at all in this limit.

Apart from this extreme case, as an example, we take the same source as in Sec.~\ref{ch:WO}, with same mass and redshift, but a lens with smaller mass, $M_{L}=100$ M$_\odot$. Even at $y=1$, with the lens being less massive, we are further into WO regime w.r.t. the previous cases. If we look at the right panel of Fig.~\ref{fig:MSD_WO}, which shows the frequency domain waveform in this regime for the $q=1$ scenario, we can see that no characteristic features are present in a very striking way, at least, not as much as in the interference regime. Nonetheless, we can see how the transformed waveforms are not simply a rescaling of the original one, with $\lambda=1$. This can be seen in the inset of the right panel of Fig.~\ref{fig:MSD_WO}. There, we plot again the lensed waveform when $\lambda=1$ and $0.5$, together with a rescaled version of the original signal, in black. The rescaling is computed in order to have the same strain at the beginning of the observation, i.e. at $f=10$ Hz. We can clearly see that, differently from the GO case, the transformation here does not introduce only a magnification. In fact, the shape of the waveform changes. This is valid also for the waveforms with other $\lambda$, and even for $\lambda=0.95$ and $1.1$.

The same, even though with more difficulties, can be observed when we move to the time domain (left panels of Fig.~\ref{fig:MSD_WO}). We can see how the most notable differences result in a rescaling of the strain magnitude. Still, some less evident features are present. For example, see the negative peak at $t\approx0.025\,s$ of red waveform on the bottom panel of Fig.~\ref{fig:MSD_WO}, which is not present when $\lambda=0.75$. This means that, even in this case, the MSD might be broken, even though it might be more difficult to check this breaking.

\section{Template matching}\label{ch:SNR}

How to quantify the precision with which the MSD could be broken in the interference regime? This is an important question to answer, because setting the range where the MSD is clearly broken means that we are able to determine where the source distance and mass lens estimation are not biased. The most direct way to assess this point would be by adopting an inference approach, for example using the inference algorithm present in \texttt{PyCBC} \cite{alex_nitz_2020_4134752}, including specific lensing routines in it. Although this is not feasible because of our hardware capabilities, we can still get important insights by calculating the matched-filtered signal-to-noise ratio (SNR) of all the cases studied above.

In a match filtering analysis, such as the one used for the detection of GWs, the SNR is calculated comparing a signal $s(t)=h(t)+n(t)$, where $h$ is the GW signal and $n$ the noise, with a template $h_{T}(t)$ \cite{maggiore2008gravitational,Ezquiaga:2020gdt}
\begin{equation}
\rho = \frac{\left(s|h_{T}\right)}{\sqrt{(h_{T}|h_{T})}} \approx \frac{\left(h|h_{T}\right)}{\sqrt{(h_{T}|h_{T})}}\, ,
\end{equation}
where the right-hand side is given if we neglect the correlation of the noise and the template. The inner product $(a|b)$ in the Fourier space is defined as
\begin{equation}
    (a|b) = 4\,{\rm Re}\left[\int_0^\infty\frac{\tilde{a}(f)\cdot\tilde{b}^*(f)}{S_n(f)}df\right]\, ,
\end{equation}
where $\tilde{a}(f)$ is the waveform (signal) in frequency domain and $S_n(f)$ is the single-sided power spectral density \cite{LIGOScientific:2019hgc,maggiore2008gravitational}.
The optimal SNR is given when the signal matches the template, $h(t)\propto h_{T}(t)$, i.e. $\rho_{\rm opt}=\sqrt{(h|h)}$.
We compute here  $\rho/\rho_{opt}$, i.e. the ratio between the SNR obtained by a given template w.r.t. the optimal SNR. To compute these calculations we use the publicly available Livingston detector sensitivity during the first months of the observing run O3\footnote{``o3\_l1" in \url{https://dcc.ligo.org/LIGO-T1500293/public}.}.

It is important to note that in order to study the distinctive features introduced by the MSD transformations we are most interested in differences in the template matching and not in the absolute SNR values. Absolute SNR values are relevant, though, to answer the question of how well the possible differences between different MSD transformations could be distinguished for a given detector network.
Since current ground-based detectors cover a similar frequency range, the ratio $\rho/\rho_{opt}$ will not change significantly among them.

In order to quantify the region of parameter space that could break the MSD, we compute the change in the $\chi^2$ when using different templates. We follow a similar approach to that of \cite{Wang:2021kzt}.
In particular, the likelihood of a given GW event can be determined assuming that after the subtraction of the waveform from the signal, the noise is Gaussian \cite{maggiore2008gravitational}, i.e.
\begin{align}
\mathcal{L} &\propto \exp \left[-\frac{1}{2}(s|s)+(h|s)-\frac{1}{2}(h|h)\right] \nonumber \\
&\propto \exp \left[(h|s)-\frac{1}{2}(h|h)\right]\, ,
\end{align}
from which the $\chi^2$ is derived to be
\begin{equation}
\chi^2 = (h|h)-2(h|s) \approx \rho^{2}_{opt} \left[1-\frac{2\rho}{\rho_{opt}}\right]\,,
\end{equation}
where we do not include the common, constant term $(s|s)$, and neglect correlations between the noise and template.
Searching for a given confidence level w.r.t. to the best model corresponding to $\rho_{opt}$ we get
\begin{equation}
\Delta \chi^2 \approx 2 \rho^{2}_{opt} \left[1-\frac{\rho}{\rho_{opt}}\right] \,.
\end{equation}
For a given threshold in $\Delta \chi^2$, this relation allows us to determine the level of mismatch, $\rho/\rho_{opt}$, that can be distinguished as a function of the optimal SNR.
When two free parameters are involved in the analysis (as in our case, as we explain in the following), the $3\sigma$ confidence level roughly corresponds to $\Delta \chi^2\sim 11.8$.
In our examples we will consider a very loud event with $\rho_{opt} \sim 55$ and a typical near threshold event with $\rho_{opt} \sim 11.5$. This will convert into $\rho/\rho_\text{opt}\approx 0.998$ and $\rho/\rho_\text{opt}\approx 0.955$ respectively. These numbers set the benchmarks to determine the region of parameter space where the MSD can be broken.
\begin{figure}[h!]
    \centering
    \includegraphics[width=0.48\textwidth]{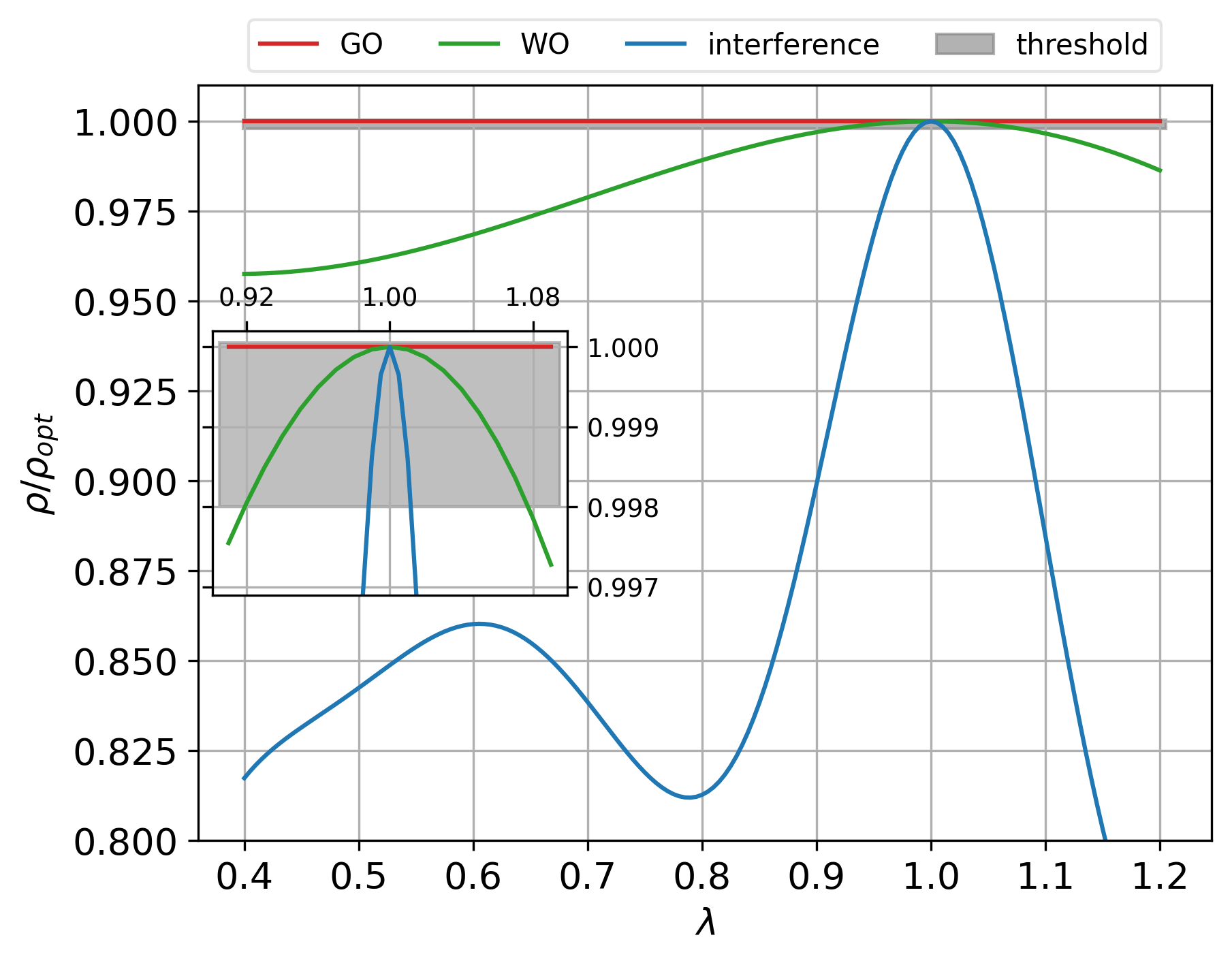}
    \caption{Effect of the MSD transformation parameter $\lambda$ on the detection of GWs. We compute the signal-to-noise ratio (SNR) using templates that vary $\lambda$, $\rho(\lambda)$, for an original signal with $\lambda=1$. We compare this SNR to the optimal SNR $\rho_{opt}=\rho(\lambda=1)$ and plot their ratio. We consider three different regimes: geometric optics (GO, red) with $M_L=500~M_\odot$ and $y=10$; interference regime (blue) with $M_L=500~M_\odot$ and $y=1$; and wave optics (WO, green) with $M_L=100~M_\odot$ and $y=1$. Lens at $z_L=0.01$, always. The source has a rest mass of $M=100~M_\odot$ at $z_S=0.1$; and $\rho_{opt} \approx 55$, which set the $3\sigma$ threshold at $0.998$.}
    \label{fig:SNR_regimes}
\end{figure}

In Fig.~\ref{fig:SNR_regimes}, we present a first insight on the behaviour of $\rho/\rho_{opt}$ depending on $\lambda$, for the three optical regimes we considered so far. The GO regimes case (in red in the figure) is given by a mass lens of $M_L=500~M_\odot$ and a source position $y=10$; the interference regime case (blue) has $M_L=500~M_\odot$ and $y=1$; and the WO case (green) has $M_L=100$ and $y=1$. The lens is always at $z_L=0.01$. The source is the same for the three cases and it is a BBH with total rest mass of $M=100~M_\odot$. The threshold is set at $0.998$ (gray region), i.e. a $3\sigma$ confidence given by $\rho_{opt}=55$.
Although this figure does not offer the full picture of the degeneracy breaking since it does not consider how $y$ and $M_L$ change with $\lambda$, it still provides valuable information of the MSD effect. In particular:
\begin{enumerate}
\item in the GO case, the MSD is not broken at all since the corresponding curve never leaves the confidence region (in fact the curve is, to our degree of numerical accuracy, flat);
\item in the interference regime, $\rho/\rho_{opt}$ decreases quickly when $\lambda\neq1$ and leaves the confidence region for $\lambda$ close to $1$, meaning that the MSD is broken for almost any $\lambda\neq1$;
\item in the WO case, $\rho/\rho_{opt}$ decrease outside $\lambda=1$, but not as quickly as in the interference case. This means that, here, the MSD is still valid for a greater interval in$~\lambda$.
\end{enumerate}

\begin{figure*}[htbp]
\centering
\includegraphics[width=0.4\textwidth]{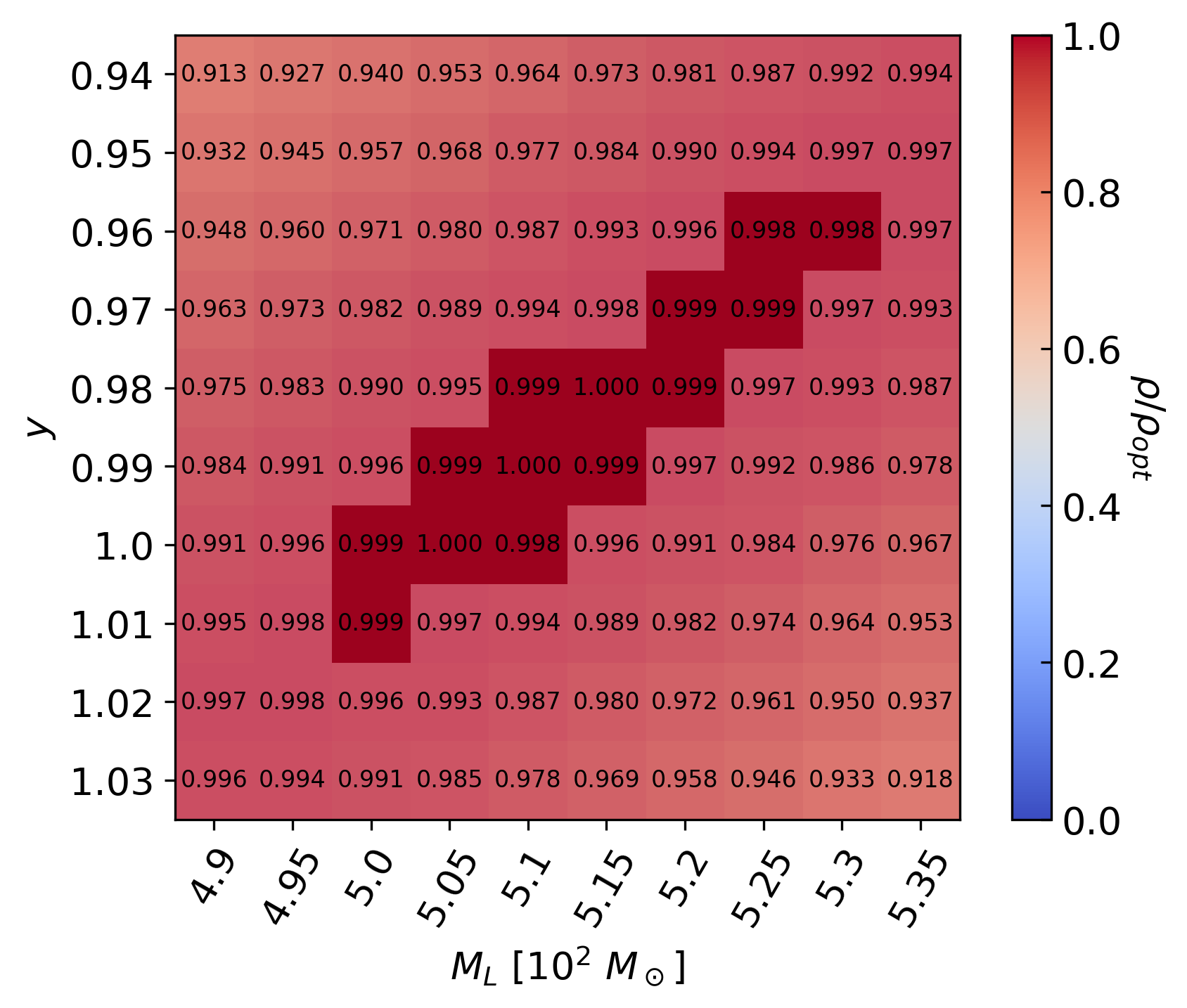}~~~
\includegraphics[width=0.4\textwidth]{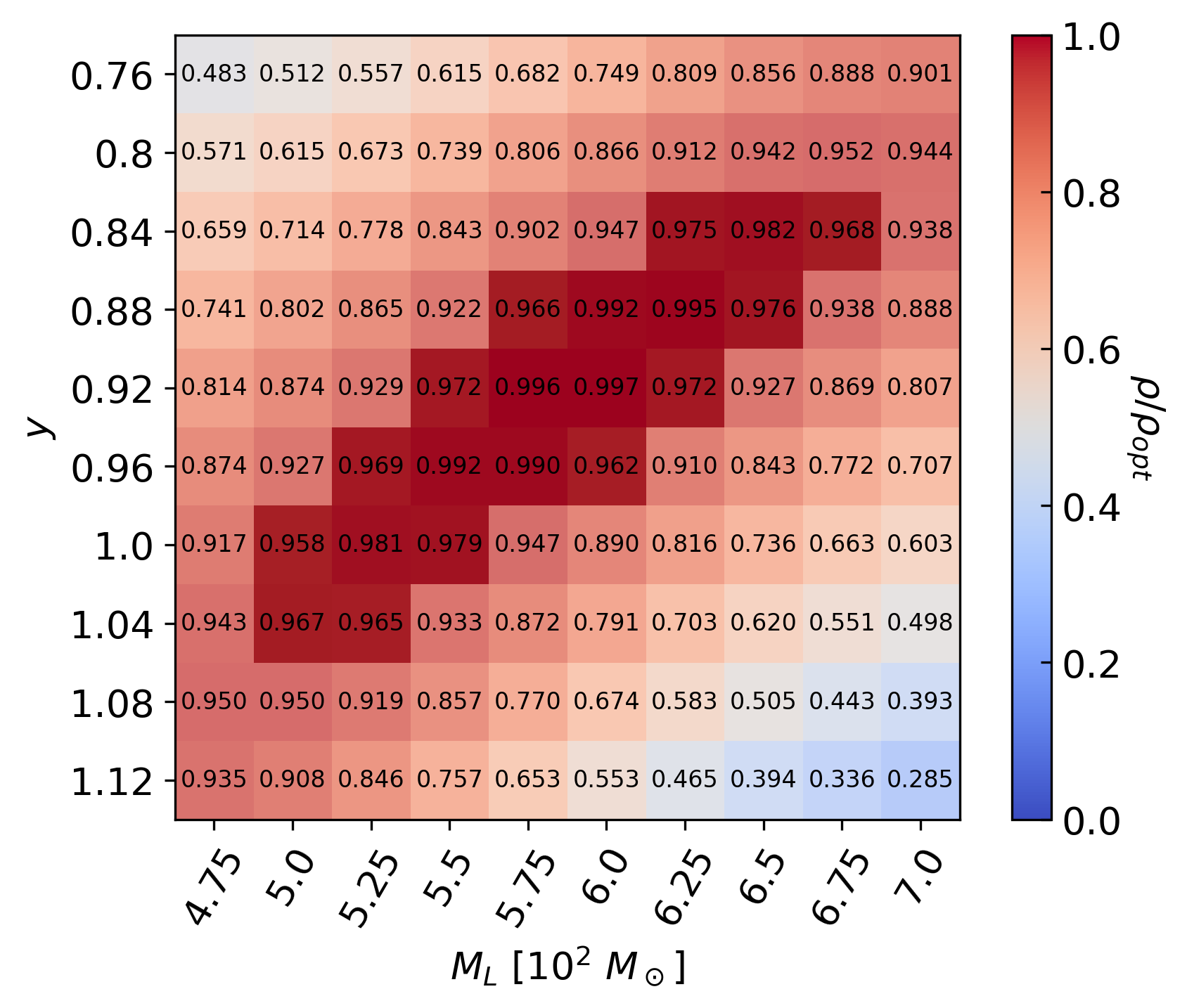}\\
\includegraphics[width=0.4\textwidth]{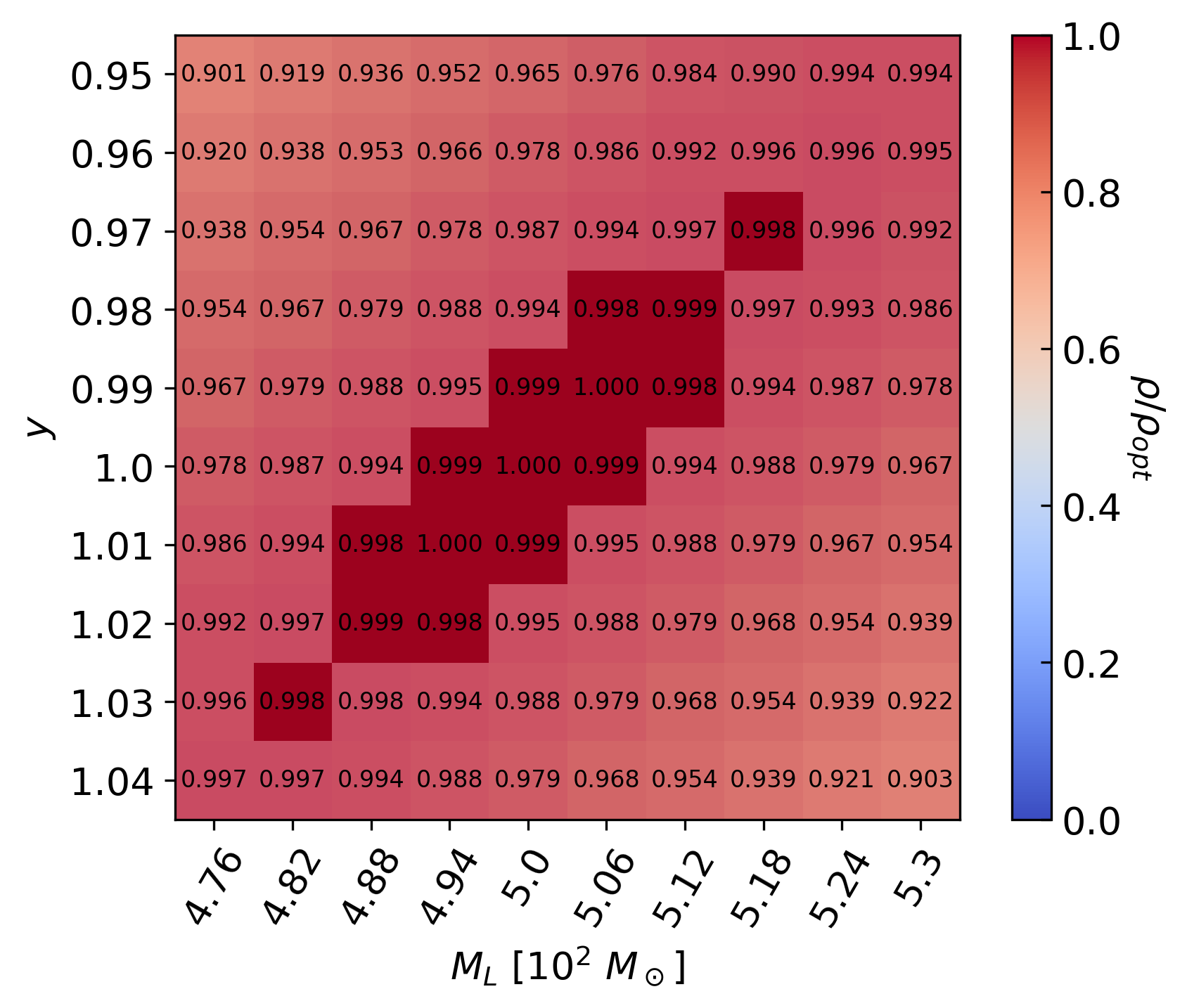}~~~
\includegraphics[width=0.4\textwidth]{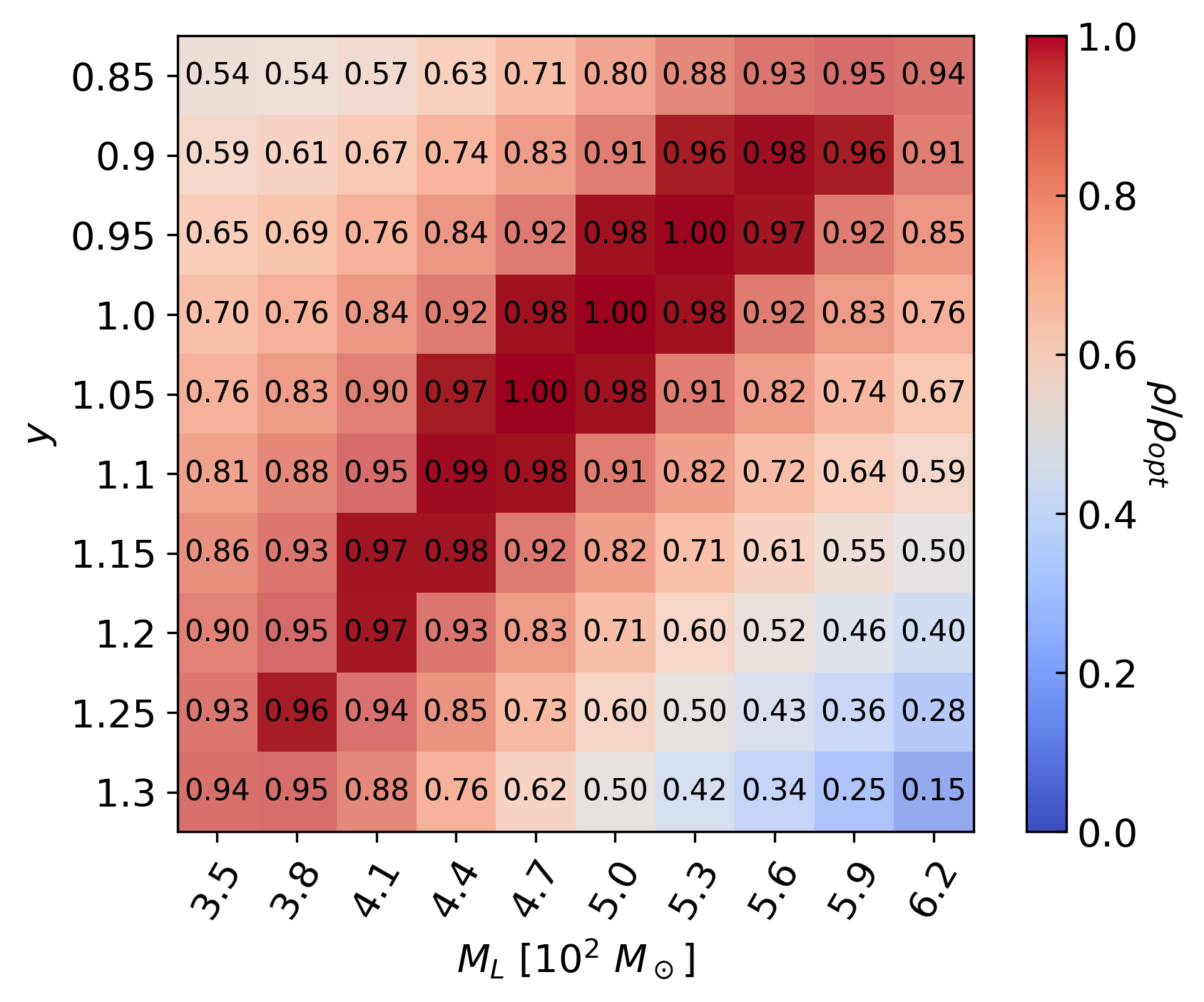}\\
\includegraphics[width=0.4\textwidth]{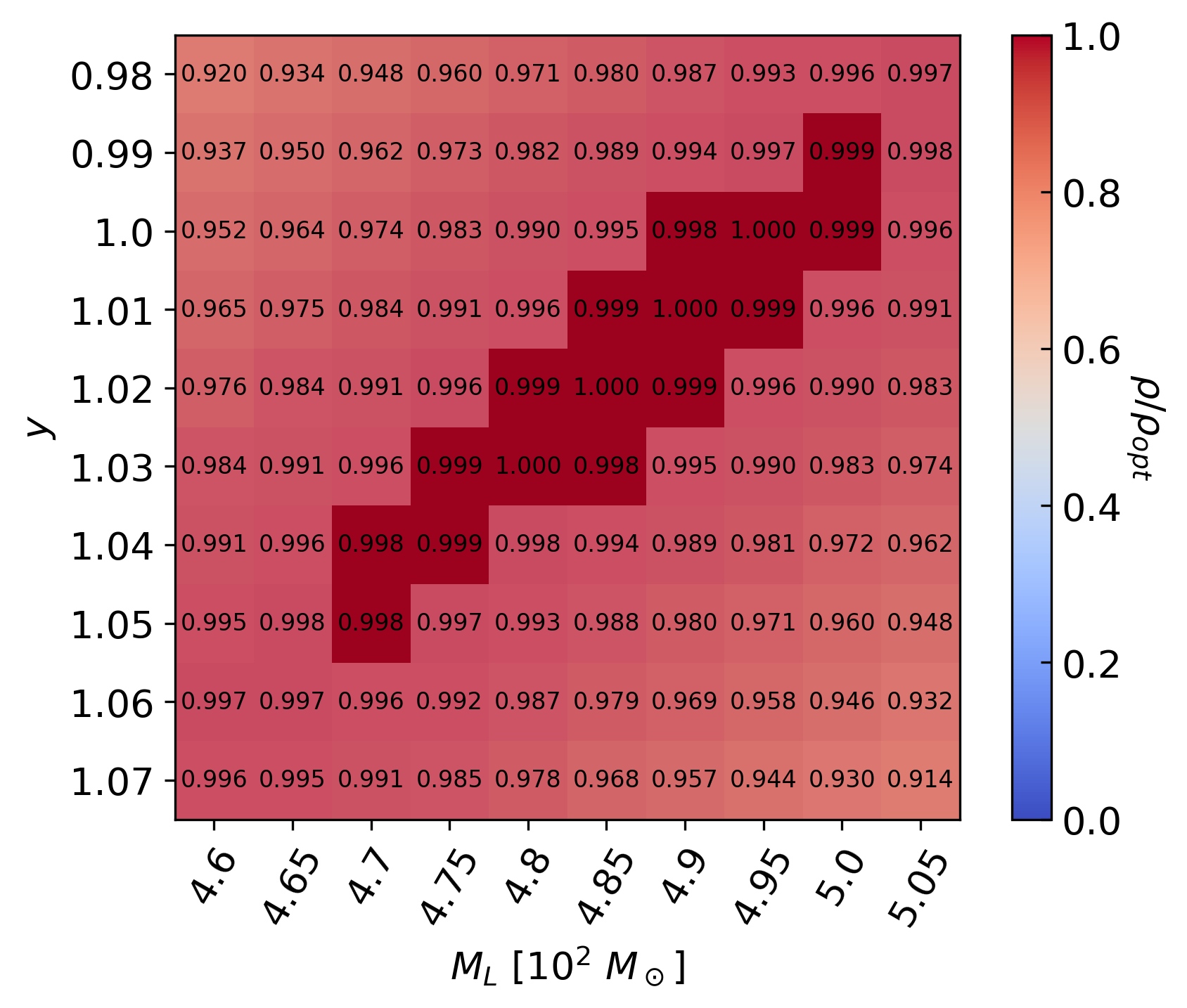}~~~
\includegraphics[width=0.4\textwidth]{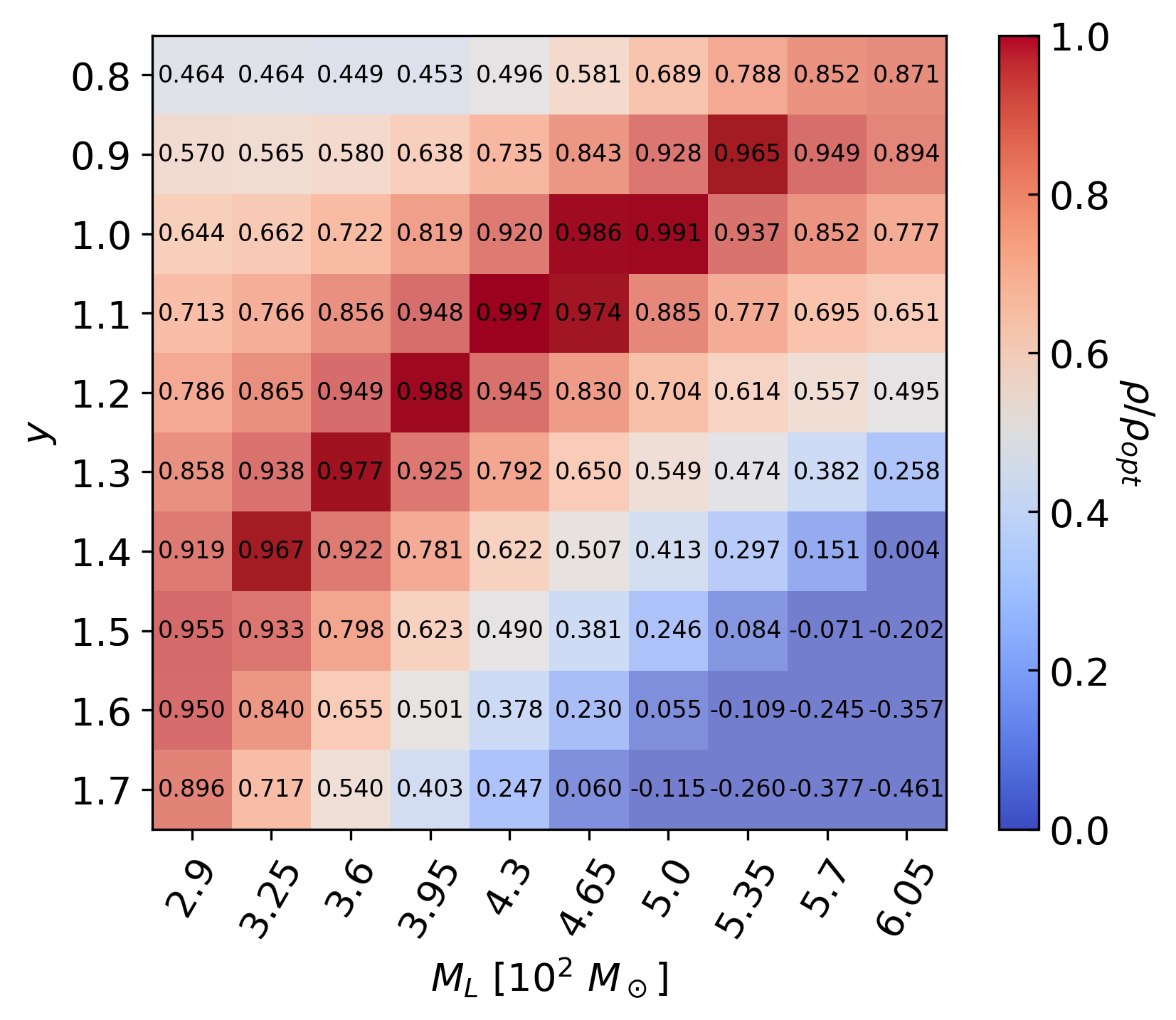}
\caption{Signal to noise ratio matrix for interference regime. \textit{(Left column.)} The signal (the same for all panels) corresponds to a source with a (rest) total mass of $M_{\rm tot} = 100~M_\odot$, $q=1$ at redshift $z_S=0.1$ and at $y=1$, and to a lens with (rest) mass $M_{L}=500~M_\odot$ at redshift $z_{L}=0.01$. The templates are calculated at the $\{y,M_L\}$ values shown in the axis, while the MSD parameter is fixed at $\lambda=\{0.99,1,1.01\}$ from top to bottom. The dark red $3\sigma$ confidence levels are calculated from $\rho_{opt}=55$ and threshold $\rho/\rho_{opt}=0.998$. \textit{(Right column.)} Same as left, but with $z_S=0.5$, and the templates calculated with the MSD parameter fixed at $\lambda=\{0.93, 1, 1.03\}$. The $3\sigma$ confidence levels are calculated from $\rho_{opt}=11.5$ and threshold $\rho/\rho_{opt}=0.955$.}
\label{fig:SNR_INT}
\end{figure*}

For a more detailed analysis, we explore the SNR mismatch in the $(M_L,y)$ plane. We present SNR calculations in form of matrices for a grid of $M_L$ and $y$.
Depending on the goal, this kind of representation: will ease the identification of the MSD range (i.e. in $\lambda$) where we can assess it is broken; will make us able to ``convert'' such $\lambda$-validity ranges into $y$ and $M_L$ ranges; and will ease the study of possible degeneracies among the parameters. We thus expect that $\rho/\rho_{opt}$ is larger than the $3\sigma$ threshold stated above not only when the template matches the signal, but also when a degeneracy is taking place. In other words, when different combinations of $y$, $M_L$ and $\lambda$ would give an indistinguishable waveform at $3\sigma$ confidence level, we may say that the MSD is on and identify the range of validity where the MSD is not broken. On the other hand, if $\rho/\rho_{opt}$ is smaller than the $3\sigma$ threshold, then we can state that, for the values of $y$ and $M_L$ for which this happens, the signal and the templates are intrinsically different, we can exclude the presence of degeneracy and we can assess that the MSD is finally broken, leading to unbiased estimation of the other parameters.

Since we have confirmed in Fig.~\ref{fig:SNR_regimes} that the MSD cannot be broken in the GO limit ($\rho(\lambda)/\rho_{opt}=1$), we focus for this analysis in the interference regime. Our results are given in Fig.~\ref{fig:SNR_INT}. The left column shows SNR calculation for this regime, when the source is at $z_S=0.1$ and has $q=1$. Each panel shows $100$ templates in correspondence of the $y$ and $M_L$ values shown on the axis. What is more important to say, is that the templates are different from one panel to the other because we consider different $\lambda$ values in each of them. In particular, $\lambda=\{0.99,1,1.01\}$, from top to bottom. The signal, instead, is always the same for each panel, and is given by $y=1$, $M_L=500~M_\odot$ and $\lambda=1$, i.e. our designed ``original'' waveform. The $3\sigma$ contours are shown as highlighted dark red boxes. We can see that when $\lambda=1$ (central left panel), we have a perfect match among templates and signal only in a very narrow region around the fiducial values corresponding to $\Delta y \sim 3\%$ and $\Delta M_L \sim 4\%$.

But if we aim at finding what could be the possible error on $y$ and $M_L$ associated to the impossibility of breaking the MSD, we must look at the panels where $\lambda = 0.99$ and $1.01$, which are the extreme cases where the $3\sigma$ confidence levels still overlap with the original signal. In that case, we can see how the most realistic $3\sigma$ uncertainty on the parameters would be $\Delta y \sim 5\%$ and $\Delta M_L \sim 6\%$.
Although we do not show other waveforms with $\lambda$ out of the range $[0.99,1.01]$, it is clear that we can now established them as safely breaking the MSD.

When we move the source further away, posing it at $z_S=0.5$, things change quantitatively, because the SNR is lower ($\rho_{opt}=11.5$), but, qualitatively speaking, we can still identify a quite narrow range under which MSD is not safely broken. From the right panel of Fig.~\ref{fig:SNR_INT} we can see how now the MSD can be considered broken for $\lambda<0.93$ and $>1.03$. The corresponding $3\sigma$ uncertainties on $y$ in this case grow up to $\Delta y \sim 16\%$ on the lower bound and $40\%$ on the upper one, while for $M_L$ we have $\Delta M_L \sim 35\%$.

Finally, we perform the same check for the WO case, which we omit to show in figures here. In this case $\rho_{opt}=61$ and the $3\sigma$ region is defined by $\rho/\rho_{opt}\geq0.998$. Here the MSD can be broken for $\lambda$ out of the range $[0.93,1.06]$, which corresponds to $3\sigma$ uncertainties of $\Delta y \sim 20\%$ and $\Delta M_L\sim 30\%$.

\section{Discussion and Conclusions}\label{ch:conclusions}

The mass-sheet degeneracy (MSD) is a well-known problem in gravitational lensing limiting the possibility of determining the lens model parameters from observations of lensed signals.
Although the implications of the MSD for lensed EM signals have been extensively studied, e.g. \cite{Falco85_MSD,Gorenstein88_MSD,Schneider:2013sxa,Schneider:2013wga,Kochanek:2019ruu,Wucknitz:2020spz,Tagore:2017pir,Blum:2020mgu}, the same is not true for lensed GWs. In this paper we have addressed this point. Understanding when the MSD can be broken is important since it will open new opportunities to infer more accurately cosmological and astrophysical parameters, e.g. H$_0$ \cite{Cremonese:2019tgb}, or lens parameters, like $M_L$ or $y$, from GWs astronomy. 

We have studied how the MSD modifies the physics of the lens-GW source system, and we have shown, both qualitatively and quantitatively, how it affects all the optics regime which can be explored in GW astronomy. In fact, contrarily to EM events, which mostly happen in the geometrical optics regime, when dealing with GWs, depending on the lens-source geometry, on the GW frequency and on the mass of the lens, we might need to rely on geometrical or wave optics formulation. Indeed, we have shown here that the most interesting scenario is related to the interference regime, intermediate between wave- and geometric-optics.

Our first result was to illustrate the effects of the MSD on the lensed waveforms in the geometric-optics approximation. We find that the usual problem connected to the MSD in the EM persists also for GW lensing without any substantial modification. In fact, the MSD only acts on the magnification of the lensed waveforms and does not alter their shape. 

The most interesting situation is related to the interference regime. Here, we demonstrate that the MSD is clearly solved because the waveform, both in the time and frequency domain, presents some characteristic features that depend on the MSD parameter, $\lambda$. In other words, the waveforms are intrinsically different if the source position and/or the lens mass are varied; thus, if the lensing analysis is included in the detection pipeline, we can infer unbiased estimation for the interested parameters.

For a full assessment of the MSD breaking it is important to determine how these signals would be heard in a given detector networks and the impact of observational errors. We have considered two different signals, both of them detectable by current LIGO/Virgo detectors, but with different SNR: one ``realistic'' case, with $\rho_{opt}=11.5$, has SNR comparable to the signals which have been detected so far; one ``optimistic'' case, with $\rho_{opt}=55$, has a SNR five times higher and can provide insights also on future GW detector sensitivities.

If we stick to the realistic case, we have found that the MSD is broken for $\lambda <0.93$ or $>1.03$. In terms of parameters which are measured, this converts in the $3\sigma$ errors $\Delta y \sim 16-40\%$ and $\Delta M_L \sim 35\%$. Thus, a rough estimation of the $1\sigma$ error would be $\Delta y \sim 5-15\%$ and $\Delta M_L \sim 12\%$. These values show that the MSD might still be a problem: in fact, realistic errors on the dispersion velocity of the lens system are of the order of $6-10\%$ \cite{Schneider:2013sxa,Kochanek:2019ruu}, which thus convert in $\Delta M_L \sim 12-20\%$.
This should be taken just as a rough estimation, considering that here we are considering point mass lenses, but it is indicative of a consistent role of the MSD in the error budget.

Instead, with higher SNRs, the MSD can be broken for $\lambda<0.99$ or $>1.01$, which implies $\Delta y \sim 5\%$ and $\Delta M_L \sim 6\%$ at $3\sigma$ or $\Delta y \sim 2\%$ and $\Delta M_L \sim 2\%$ at $1\sigma$. Thus, in this case, the role of the MSD would be clearly subdominant with respect to other known sources of uncertainty.

Finally, we studied the MSD in the wave-optics regime. We have demonstrated that, contrary to the geometric-optics case and similarly to the interference one, the MSD does not simply rescale the waveform, but intrinsically changes its shape, making it possible to break the degeneracy. However, the lensing effects being much weaker than in the interference regime, the MSD breaking does not perform as well as in the interference regime.

We plan to extend our analysis in several ways. First of all, one could improve the quantitative part, possibly through direct inference, on how well the MSD can be solved in all regimes. Then, we could expand the study to different lens models, while here we were restricted to a point mass only. One could also expand the study to different sources, considering BBHs with different masses and/or at higher redshifts, and possibly considering also binary neutron stars. And, finally, a deeper analysis of the WO regime should be performed, in order to define a better limit up to which the MSD can be solved.

\section*{Acknowledgments}

P.C. is supported by the project ``Uniwersytet 2.0 - Strefa Kariery, ``Miedzynarodowe Studia Doktoranckie Nauk Scislych WMF”, nr POWR.03.05.00-00-Z064/17-00'' co-funded by the European Union through the European Social Fund.

J.M.E. is supported by NASA through the NASA Hubble Fellowship grant HST-HF2-51435.001-A awarded by the Space Telescope Science Institute, which is operated by the Association of Universities for Research in Astronomy, Inc., for NASA, under contract NAS5-26555. He is also supported by the Kavli Institute for Cosmological Physics through an endowment from the Kavli Foundation and its founder Fred Kavli.

\bibliographystyle{apsrev4-1}
\bibliography{biblio}{}

\end{document}